\documentclass[eprint,superscriptaddress,reprint]{revtex4-2}
\usepackage{float}
\usepackage{graphicx}
\usepackage{xcolor}
\usepackage{amsmath}
\usepackage{multirow}

\usepackage{color} 
\usepackage{comment} 
\usepackage{braket} 
\usepackage{esint} 
\usepackage{amsfonts}
\usepackage{amssymb}
\usepackage{bm}
\usepackage{dcolumn}

\usepackage[colorlinks,linkcolor=blue,anchorcolor=blue,urlcolor=blue,urlcolor=blue,citecolor=blue]{hyperref}

\usepackage{soul}

\begin{document}

\title{Extensible Fluxonium Architecture Using Tunable Couplers \\ with Low Shunt Capacitance}

\author{Peng Zhao}
\email{shangniguo@sina.com}
\affiliation{Quantum Science Center of Guangdong-Hong Kong-Macao Greater Bay Area, Shenzhen 518045, China}
\author{Peng Xu}
\affiliation{Institute of Quantum Information and Technology,
Nanjing University of Posts and Telecommunications, Nanjing, Jiangsu 210003, China}
\affiliation{Jiangsu Key Laboratory of Quantum Information Science and Technology, Nanjing University, Suzhou, 215163 Jiangsu, China}
\author{Zheng-Yuan Xue}
\email{zyxue83@163.com}
\affiliation{Key Laboratory of Atomic and Subatomic Structure and Quantum Control (South China Normal University), Ministry of Education, Guangdong Basic Research Center of Excellence for Structure and Fundamental Interactions of Matter, and School of Physics, South China Normal University, Guangzhou 510006, China}
\affiliation{Guangdong Provincial Key Laboratory of Quantum Engineering and Quantum Materials, Guangdong-Hong Kong Joint Laboratory of Quantum Matter, and Frontier Research Institute for Physics, South China Normal University, Guangzhou 510006, China}
\affiliation{Quantum Science Center of Guangdong-Hong Kong-Macao Greater Bay Area, Shenzhen 518045, China}

\date{\today}

\begin{abstract}

Fluxonium qubits have demonstrated high-fidelity operations and long coherence times in small-scale systems, highlighting their promise for quantum computing. However, large-scale 
integration into a high-performance two-dimensional (2D) qubit array remains the central challenge 
for practical applications. In this work, we introduce an extensible architecture for scaling up 
fluxonium qubits in 2D grids. To address the key challenges, namely achieving controllable strong interaction and high connectivity for qubits featuring small shunting capacitors (footprints), we propose using 
low-shunt-capacitance couplers to enable tunable interactions between fluxonium qubits. When embedded into 2D square lattices, large couplings can be achieved even with relatively small coupling capacitances, thus 
enabling multiple connections and other applications with sufficient capacitance budget. We 
further propose coupler realizations based on generalized flux 
qubit circuits, specifically the quarton and the fluxonium, and demonstrate that both 
enable fast, high-fidelity gates with low spectator errors, while supporting multiple 
connections on 2D grids.

\end{abstract}

\maketitle


\section{Introduction}\label{SecI}

Fluxonium qubits, with long coherence times and strong anharmonicity, have emerged as a promising alternative for superconducting quantum processors~\cite{Jiang2025,Manucharyan2009,Nguyen2019}.
Significant progress has been made in improving performance, with qubit lifetimes now extending into the millisecond regime~\cite{Somoroff2023} and gate fidelities well above $99.9\%$~\cite{Somoroff2023,Rower2024,Ding2023,Zhang2024,Lin2025}. Nevertheless, these demonstrations have been limited to small-scale processors, typically comprising only two qubits, and remain far from the scale required for practical quantum utility, let alone fault-tolerant quantum computing. Therefore, while continued efforts are needed to further boost small-scale performance, attention must also turn to scaling fluxonium qubits to large-scale, two-dimensional (2D) architectures, if they are to compete with conventional transmon qubits~\cite{Koch2007}.

In scaling fluxonium qubits from one to two and eventually to many qubits arranged in 2D 
lattices, the critical prerequisite is to realize scalable coupling schemes that, from a 
single-qubit perspective, handle issues from growing spectator qubits. As with transmon qubits, fluxonium qubits 
connected via always-on couplings, whether direct or indirect, suffer from significant 
spectator-induced crosstalk that substantially degrades performance~\cite{Zhao2026,Zwanenburg2026,Zhao2026a,Zhan2026,Chan2026}. Although careful frequency 
allocation may mitigate this issue~\cite{Nesterov2018,Nguyen2022,Moreno2025,Kugut2025,Huang2026}, such 
strategies impose significant fabrication burdens and offer limited flexibility against unexpected frequency 
collisions~\cite{Brink2018,Malekakhlagh2020,Zhao2023,Heya2024}. This motivates the adoption of tunable couplers~\cite{Zhang2024,Zhao2026,Zhao2026a,Zwanenburg2026,Zhan2026,Chan2026,Moskalenko2021,Moskalenko2022}, such 
as the inductive coupler~\cite{Chen2014}, the transmon coupler~\cite{Yan2018}, and the double-transmon coupler~\cite{Goto2022,Campbell2023} (the ultrastrong inter-resonator coupler~\cite{Miyanaga2021}), to address spectator-induced errors in fluxonium qubits. Beyond this, a pivotal challenge in scaling fluxonium 
qubits is to realize the chosen coupler design in genuinely extensible 2D lattices. Unlike transmon qubits, which have large shunt capacitors (large footprints), fluxonium qubits typically rely on small shunt capacitances (and thus small footprints). This makes accommodating the multiple connections needed in 2D lattices particularly challenging, especially given the large coupling capacitances needed for strong coupling and fast gates~\cite{Zhao2026,Zhao2026a,Zwanenburg2026,Zhan2026,Chan2026,Zhao2026b}. Thus, although previous designs have proven effective at eliminating spectator-induced errors and enable fast gates~\cite{Zhang2024,Zhan2026,Moskalenko2022}, embedding them into 2D grids remains a nontrivial challenge.

To move fluxonium qubits beyond few-qubit prototypes toward truly large-scale 2D lattices, we 
here propose an extensible coupling architecture based on low-shunt-capacitance tunable couplers. 
The coupler design relies on the generalized flux-qubit circuit, which features a much smaller shunt 
capacitance. As a result, strong and controllable interactions can be realized with relatively 
small coupling capacitances, leaving a sufficient capacitance budget for multiple connections 
in 2D lattices. As concrete realizations, we consider two specific coupler designs and demonstrate that both can achieve fast, high-fidelity two-qubit gates with low spectator errors. 

\section{Low-shunt-capacitance design}\label{SecII}

\begin{figure}[tbp]
\begin{center}
\includegraphics[keepaspectratio=true,width=\columnwidth]{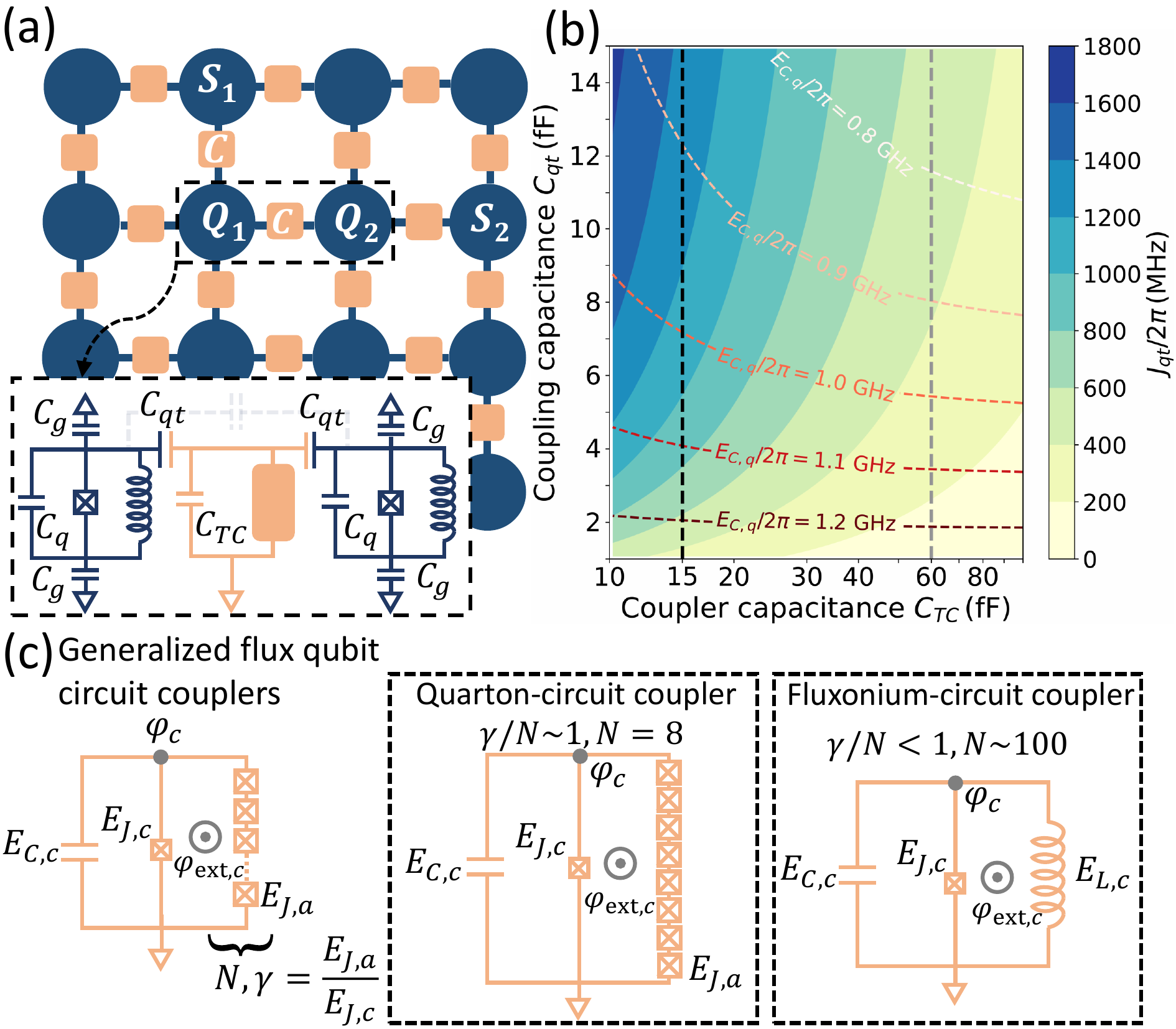}
\end{center}
\caption{(a) A 2D square lattice comprising fluxonium qubits (circles) coupled via couplers (squares). 
The inset depicts a tunable capacitive coupling architecture, where qubits are capacitively 
coupled ($C_{qt}$) to a tunable coupler with shunt capacitance $C_{TC}$. (b) Coupling energies $J_{qt}$ (filled contours) and qubit charging energies $E_{C,q}$ (dashed contours) 
as functions of the coupling capacitance $C_{qt}$ and the coupler capacitance $C_{TC}$ for the unit 
cell. The unit cell consists of two floating fluxonium qubits [$C_{q\,(g)}=7\,(13)\,\rm fF$] coupled via 
a grounded coupler, embedded in the 2D lattice shown in (a). The vertical dashed grey and black lines 
mark, respectively, the typical shunt capacitances for transmon-based coupler designs and for 
the proposed design based on the generalized flux-qubit circuit shown in (c). This circuit comprises 
a Josephson junction ($E_{J,c}$) shunted by both a small capacitor ($E_{C,c}$) and an $N$-junction array ($E_{J,a}$). 
Depending on $N$ and $\gamma=E_{J,a}/E_{J,c}$, the circuit behaves as a quarton when $\gamma/N\sim 1$ (e.g., $N=8$), and as a fluxonium when $\gamma/N< 1$ (e.g., $N\sim 100$).}
\label{fig1}
\end{figure}

Fluxonium qubits, with transition frequencies of $0.1-1\,\rm GHz$~\cite{Nguyen2019,Nguyen2022}, feature a small shunt capacitance (small footprint), corresponding to a typical charging energy $\sim 1\,\rm GHz$ (a total capacitance $\sim 20\,\rm fF$)~\cite{Nguyen2022}. Consequently, the coupling capacitances between a fluxonium qubit and other elements, such as couplers and readout resonators, can contribute significantly to its charging energy. Moreover, achieving large coupling capacitances while avoiding spurious coupling is physically challenging for small-footprint qubits in actual circuit layouts, especially given the need for multiple connections in 2D grids~\cite{Ding2023,Zhan2026}. All these factors severely constrain both qubit layout and coupling design, a challenge we refer to as the capacitance loading issue~\cite{Ding2023,Zhao2026b,Rosenfeld2024}, which is absent in conventional transmons, given their large shunt capacitors (large footprint). Before detailing our design to address this issue, we first illustrate it more explicitly using the generalized circuit model shown in Fig.~\ref{fig1}(a).

Without loss of generality, we consider the unit cell, where two floating fluxonium qubits (with 
island capacitance $C_{g}$ and shunt capacitance $C_{q}$) are capacitively 
coupled (coupling capacitance $C_{qt}$) to a coupler with shunt capacitance $C_{TC}$, embedded 
in the 2D lattice. When $C_{qt}\ll \{C_{q},C_{g},C_{TC}\}$, the fluxonium-coupler coupling 
energy is, to leading order, 
\begin{equation}
\begin{aligned}\label{eq1}
J_{qt}\propto \frac{C_{qt}}{C_{TC}}
\end{aligned}
\end{equation}
(also holds for grounded qubits; see Appendix~\ref{A}). Thus, for a given 
$J_{qt}$, low-shunt-capacitance couplers require smaller coupling capacitances 
than their large-shunt-capacitance counterparts.

This observation holds even when $C_{qt}$ is comparable to 
$\{C_{q},C_{g},C_{TC}\}$, as illustrated in Fig.~\ref{fig1}(b) for the concrete case 
$C_{q\,(g)}=7\,(13)\,\rm fF$. For the shunt capacitance 
of $15\,\rm fF$ (vertical black dashed line), the coupler design allows the coupling energy $J_{qt}$ to reach $1\,\rm GHz$ (with $C_{qt}\approx 5.5\,\rm fF$), while maintaining a qubit charging energy $E_{C,q}/2\pi\approx 1.047\,\rm GHz$, well within the typical fluxonium regime. In contrast, for transmon-based designs~\cite{Ding2023,Zhao2026,Zhao2026a,Zwanenburg2026,Zhan2026,Chan2026} with typical $C_{TC}=60\,\rm fF$ (vertical dashed grey line), achieving such strong coupling while keeping the qubit within the fluxonium regime is impractical.

We therefore favor low-shunt-capacitance designs to address the capacitance 
loading issue. As shown in Fig.~\ref{fig1}(c), we consider coupler designs based on the generalized flux-qubit circuit, in which a main Josephson junction (with Josephson energy $E_{J,c}$) is shunted by a small capacitor (with charging energy $E_{C,c}$) and an array of $N$ Josephson junctions (each with Josephson energy $E_{J,a}$)~\cite{Yan2020}. The circuit dynamics can be described by two conjugate variables: the phase $\varphi_c$ across the main junction and the Cooper-pair number $n_{c}$. The potential profile is determined by $N$ and $\gamma=E_{J,a}/E_{J,c}$, defining various circuit types in distinct parameter regimes. We focus on two particular regimes, corresponding to the quarton circuit with $\gamma/N\sim 1$ (e.g., $N=8$)~\cite{Yan2020} and the fluxonium circuit with $\gamma/N< 1$ (e.g., $N\sim 100$, where the junction array acts as an inductor characterized by the inductive energy $E_{L,c}$~\cite{Nguyen2019,Nguyen2022}). For both designs, the shunt capacitance can typically be as low as $15\,\rm fF$ (total capacitance $\sim 20\,\rm fF$ and the charging energy $E_{C,c}/2\pi\sim 1\,\rm GHz$)~\cite{Yan2020,Nguyen2022}. The two coupler circuits can be approximated by the following Hamiltonians~\cite{Manucharyan2009,Yan2020}:
\begin{equation}
\begin{aligned}\label{eq2}
&\hat{H}_{Q}= 4 E_{C,c} \hat{n}^2_c-E_{J,c}\left[\gamma N \cos\left(\frac{\hat\varphi_c}{N}\right)+\cos(\hat\varphi_c -\varphi_{\text{ext},c})\right]
\end{aligned}
\end{equation}
for the quarton circuit and 
\begin{equation}
\begin{aligned}\label{eq3}
\hat{H}_{F}= 4 E_{C,c} \hat{n}^2_c+\frac{E_{L,c}}{2}(\hat\varphi_c - \varphi_{\text{ext},c})^2-E_{J,c}\cos\hat\varphi_c,
\end{aligned}
\end{equation}
for the fluxonium circuit, where $\varphi_{\text{ext},c}=2\pi\Phi_{\text{ext},c}/\Phi_0$ is the external flux bias and $\Phi_0$ is the flux quantum.  

\begin{figure}[tbp]
\begin{center}
\includegraphics[keepaspectratio=true,width=\columnwidth]{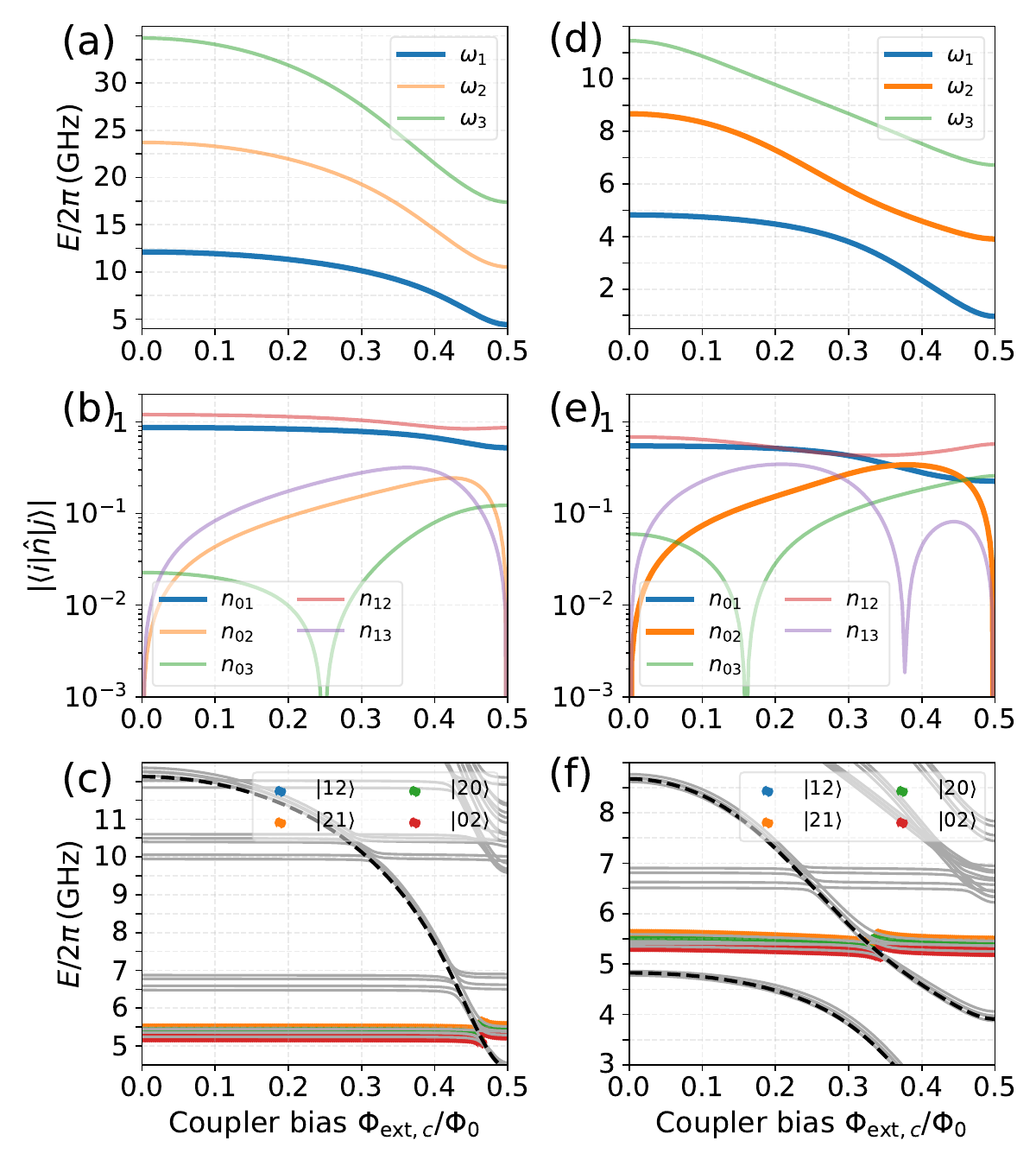}
\end{center}
\caption{(a) Energy levels of the quarton coupler, (b) magnitudes of the electric transition dipole moments for the quarton transitions, and (c) full energy levels of the coupled system consisting of two fluxonium qubits coupled via the quarton coupler, versus coupler flux bias. The noncomputational energy levels $\{|20\rangle,|02\rangle,|21\rangle,|12\rangle\}$ associated with the plasmon transition $|1\rangle \leftrightarrow |2\rangle$, together with the bare coupler frequency [black dashed lines, as in (a)], are highlighted. (d)–(f) Corresponding results for the fluxonium-circuit coupler.}
\label{fig2}
\end{figure}

As shown in Fig.~\ref{fig1}(b) and Eq.~(\ref{eq1}), for a given coupling energy $J_{qt}$, the proposed design reduces the required coupling capacitance $C_{qt}$ by a factor of $\sim 4$ in the small-coupling limit, and by an even larger factor outside this regime, compared to conventional transmon-based designs. However, gate speed is more directly determined by the coupling strength $g_{qt}\propto J_{qt} n_{01}$ than by the coupling energy $J_{qt}$ alone. Here, $n_{01}$ is the charge dipole moment for the coupler transition $|0\rangle\leftrightarrow|1\rangle$. Typically, $n_{01}\sim 1$ for transmon qubits~\cite{Koch2007}, whereas for quarton and fluxonium couplers $n_{01}\sim 0.5$ [see Figs.~\ref{fig2}(b) and~\ref{fig2}(e)]. For a given coupling strength (gate speed), the smaller dipole thus partially offsets this advantage. Nevertheless, the required $C_{qt}$ is still reduced by a factor of approximately 2 in the small-coupling limit, and more substantially beyond it, compared to transmon-based designs. 

Concretely, to achieve the coupling strength of $250\,(500)\,\rm MHz$, the proposed design requires $C_{qt}\approx 1.7\, (5.5)\,\rm fF$, whereas for the transmon-based designs, $C_{qt}\approx 3.5\,(12.2)\,\rm fF$. Moreover, for large coupling strengths, fluxonium qubits with the proposed couplers remain well within the fluxonium regime, whereas those with transmon-based designs fall outside it. We thus expect that such designs can support multiple connections in 2D grids.

\begin{table}[!htb]
\caption{\label{tab:circuit_parameters} Hamiltonian parameters of the coupled
circuit.}
\begin{ruledtabular}
\begin{tabular}{cccc}
(GHz) & $E_C/2\pi$& $E_L/2\pi$& $E_J/2\pi$ \\\hline
Fluxonium $Q_{1}$ & 1.00 & 0.80 & 6.40  \\
Fluxonium $Q_{2}$ & 1.00 & 0.60 & 6.00 \\
Spectator $S_{1}$ & 1.00 & 0.61 & 6.05 \\
Spectator $S_{2}$ & 1.00 & 0.81 & 6.45 \\
Quarton coupler $C_{Q}$   & 1.00 & ($\gamma/N=0.92$, $N=8$) & 10.43\\
Fluxonium coupler $C_{F}$ & 1.00 & 1.00 & 3.00 \\
\hline
\hline
(MHz) & $J_{c1}/2\pi$  & $J_{c2}/2\pi$  & $J_{12}/2\pi$ \\\hline
Coupling ($C_{Q}$)  & 550 & 550 & 47 \\
Coupling ($C_{F}$) & 650  & 650  & 10 
\end{tabular}
\end{ruledtabular}
\end{table}

\section{Tunable plasmon interaction and spectator error suppression}\label{SecIII}

\begin{figure*}[tbp]
\begin{center}
\includegraphics[width=16cm,height=7cm]{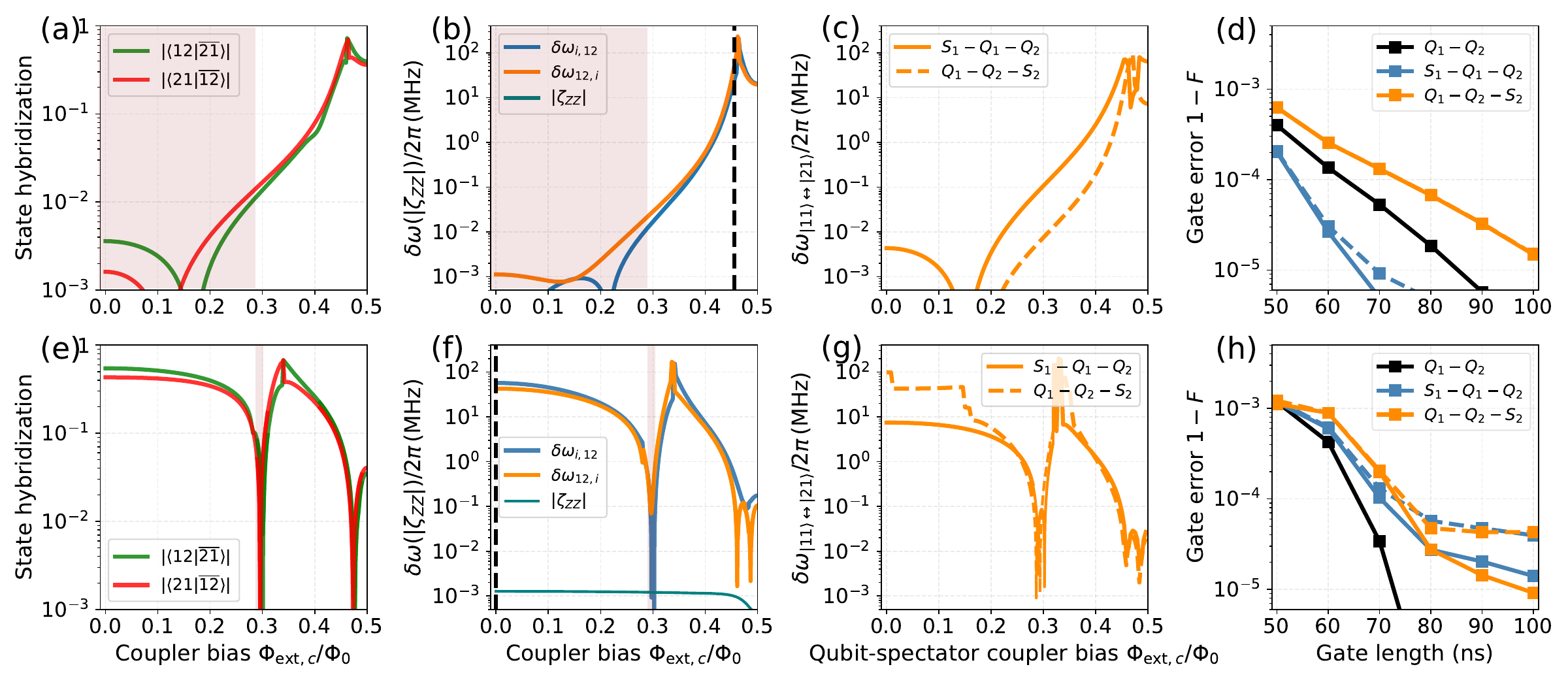}
\end{center}
\caption{(a) State hybridization and (b) state-dependent frequency shift of $|1\rangle\leftrightarrow|2\rangle$ transition versus coupler bias, for 
fluxonium qubits $Q_{1}$ and $Q_{2}$ coupled via the quarton-circuit coupler. The shaded region 
marks the bias range where plasmon interactions are strongly suppressed. (c) Frequency 
shift in the gate transition $|11\rangle\leftrightarrow|21\rangle$ induced by residual spectator-qubit 
coupling, controlled by the flux bias applied to the qubit-spectator coupler. The coupler bias 
for $Q_{1}$ and $Q_{2}$ is indicated by the vertical dashed black line in (b). Solid and dashed 
lines correspond to the configurations $S_{1}-Q_{1}-Q_{2}$ and $Q_{1}-Q_{2}-S_{2}$, respectively [see Fig.~\ref{fig1}(a)]. (d) CZ gate errors for $Q_{1}$ and $Q_{2}$ versus gate length, with 
spectator $S_{1}$ or $S_{2}$ present. Solid and dashed lines correspond to the results with the spectator 
prepared in $|0\rangle$ and $|1\rangle$, respectively. The error for the isolated $Q_{1}-Q_{2}$ system is shown for comparison. In (a)-(c), abrupt jumps originate from state-labeling failures near avoided crossings. (e)-(h) Corresponding results for the fluxonium-circuit coupler.}
\label{fig3}
\end{figure*}

To ensure extensibility to 2D lattices, we must also 
examine how such designs can address spectator-induced errors. Before 
presenting the details, we recall a key feature of fluxonium qubits biased 
at the half-integer flux quantum ($\{|0\rangle,|1\rangle,|2\rangle,|3\rangle\}$ denoting the lowest four levels). The dipole moment of the qubit transition $|0\rangle\leftrightarrow|1\rangle$ is rather weak, so when qubits are capacitively coupled via a coupler, the qubit states remain nearly decoupled from the coupler~\cite{Ding2023,Zhao2026}. 
This provides passive suppression of both the coupler-mediated $XY$ and $ZZ$ interactions between qubits~\cite{Ding2023,Zhao2026,Zhan2026}. By contrast, the dipole moments for transitions to higher-energy levels, such as $|1\rangle\leftrightarrow|2\rangle$ (referred to as plasmon transitions hereafter), are comparable 
to those of transmons (see Fig.~\ref{fig2}). The coupler can therefore interact strongly to these plasmon transitions and mediate substantial interactions between them~\cite{Ding2023,Zhao2026,Rosenfeld2024}, hereafter referred to as plasmon interactions. While plasmon interactions enable entangling gates~\cite{Ding2023,Nesterov2018,Ficheux2021,Simakov2023,Zhao2025c} such as microwave-activated C-phase (MAP) gates~\cite{Nesterov2018,Chow2013}, the lack of tunable control makes them impractical for large-scale systems due to spectator errors from always-on interactions~\cite{Zhao2026,Zhan2026,Zwanenburg2026}. We therefore adopt the proposed coupler designs to realize tunable plasmon interactions, focusing mainly on the $|1\rangle\leftrightarrow|2\rangle$ transition.

For fluxonium qubits $Q_{1}$ and $Q_{2}$ coupled via the proposed couplers [see Fig.~\ref{fig1}(c)], the full system Hamiltonian is 
\begin{equation}
\begin{aligned}\label{eq4}
\hat{H}= &\sum_{j=1,2} [\hat{H}_{fj}+J_{cj} \hat{n}_{fj}\hat{n}_{c}]+\hat{H}_{c}+J_{12}\hat{n}_{f1} \hat{n}_{f2},
\end{aligned}
\end{equation}
where subscripts~${j=1,2}$ and~$c$ refer to the fluxonium qubits and the coupler, respectively. 
Here, $\hat{H}_{fj}$ is the Hamiltonian for $Q_{j}$, of the same form as the fluxonium-coupler Hamiltonian in Eq.~(\ref{eq3}), and $\hat{H}_{c}$ denotes the coupler Hamiltonian, given for the quarton and fluxonium couplers in Eqs.~(\ref{eq2}) and~(\ref{eq3}), respectively. $J_{cj}$ denotes the qubit–coupler coupling energy and $J_{12}$ denotes the direct qubit-qubit coupling energy (see Appendix~\ref{A}). Hereafter, the system state is denoted as $|Q_{1}Q_{2}\rangle$. Using the parameters listed in Table~\ref{tab:circuit_parameters}, Figure~\ref{fig2} shows the energy levels and electric transition dipole moments of both coupler designs, together with the 
spectrum of the full coupled system described by Eq.~(\ref{eq4}), versus coupler bias. 

For the quarton-circuit coupler, it behaves like a conventional transmon: its transition frequency is flux-tunable [see Fig.~\ref{fig2}(a)], and the dipole moments for transitions between neighboring states dominate over those for non-neighboring transitions across the full bias range [see Fig.~\ref{fig2}(b)]. Moreover, its $|0\rangle\leftrightarrow|1\rangle$ transition can be tuned from $\sim 10\,\rm GHz$ 
to $\sim 5\,\rm GHz$, well within the typical frequency range of the fluxonium $|1\rangle\leftrightarrow|2\rangle$ transition. This enables strong interactions between the quarton and the fluxonium plasmon transition, as evidenced by the large avoided crossings shown in Fig.~\ref{fig2}(c). Meanwhile, as 
noted earlier, the computational subspace remains nearly decoupled from the coupler due 
to its weak transition dipole, which is confirmed by the almost flux-independent 
qubit spectrum (see Appendix~\ref{B}).

Thus, by combining the coupler-mediated coupling, which scales as $(J_{cj}|n_{fj,12}n_{c,01}|)^{2}/\Delta_{qc}$ with $\Delta_{qc}$ the detuning between the plasmon transition and the coupler, and the direct, fixed qubit coupling $J_{12}|n_{f1,12}n_{f2,12}|$, the plasmon interaction ($|12\rangle\leftrightarrow|21\rangle$) can 
be turned on and off simply by tuning the quarton frequency (hence $\Delta_{qc}$, as in Refs.~\cite{Yan2018,Niskanen2007,Mundada2019}, see Appendix~\ref{C}). This tunability is illustrated in Figs.~\ref{fig3}(a) and~\ref{fig3}(b), where the plasmon interaction is quantified by the state hybridization and the state-dependent frequency shifts~\cite{Zhao2026}. The state hybridization is defined as the overlap 
between the system eigenstates $|kl\rangle$ and the corresponding bare states $|\overline{lk}\rangle$, and the state-dependent frequency shifts are $\delta\omega_{12,i}=|(E_{21}-E_{11})-(E_{20}-E_{10})|$ and $\delta\omega_{i,12}=|(E_{12}-E_{11})-(E_{02}-E_{01})|$, with $E_{kl}$ denoting the energy of eigenstate 
$|kl\rangle$. As expected, when the coupler is far detuned from the plasmon transition, the interaction is turned off (highlighted by the shaded region), with both the hybridization and frequency shifts strongly suppressed. As the coupler is tuned closer to the plasmon transition, state-dependent frequency shifts of tens of MHz become achievable, enabling sub‑100 ns CZ gates via the MAP scheme (see Appendix~\ref{D}). The $ZZ$ coupling, $\zeta_{ZZ}=(E_{11}-E_{01})-(E_{10}-E_{00})$~\cite{DiCarlo2009}, is also shown and remains below $1\,\rm kHz$ across the entire bias range, confirming the decoupling of the computational states mentioned earlier.

To assess the efficiency of this interaction tunability in addressing spectator errors, we consider two configurations, $S_{1}-Q_{1}-Q_{2}$ and $Q_{1}-Q_{2}-S_{2}$ (see Fig.~\ref{fig1}(a)), and examine the effect of the spectator 
qubit ($S_{1}$ and $S_{2}$) on CZ gates for $Q_{1}$ and $ Q_{2}$~\cite{Zhao2026,Krinner2020,Cai2021}. The gate is realized by selectively 
activating the $|11\rangle\leftrightarrow|21\rangle$ transition following the MAP gate 
scheme with cosine DRAG pulses~\cite{Motzoi2009} (see Appendix~\ref{D}). Figure~\ref{fig3}(c) shows the frequency shift in the gate transition (the difference between the gate transition frequencies with the spectator in $|0\rangle$ and $|1\rangle$) induced by the spectator-qubit coupling (tuned by the qubit–spectator coupler flux bias). When a large residual coupling is present, this shift can reach tens of MHz, significantly affecting CZ gate performance, and could in fact be exploited to implement native CCZ gates~\cite{Zhao2026a}. When the qubit–spectator coupling is turned off, the spectator-induced frequency shift is suppressed below $1\,\rm kHz$, leading to negligible spectator-state dependence of the gate error (below $10^{-5}$)~\cite{Pedersen2007}, as shown in Fig.~\ref{fig3}(d). Moreover, the gate fidelity in the presence of a spectator remains comparable to that of an isolated two-qubit system.

Here we turn to the fluxonium-circuit coupler. As shown in Figs.~\ref{fig2}(d) and~\ref{fig2}(e), when biased away from the half-integer flux quantum, the fluxonium circuit's $|0\rangle\leftrightarrow|1\rangle$ transition is flux-tunable and falls well within the frequency range of the qubit's $|1\rangle\leftrightarrow|2\rangle$ transition. Although the coupler transition $|0\rangle\leftrightarrow|2\rangle$ is not forbidden away from the two sweet spots, its frequency is significantly higher and its dipole moment weaker until the coupler is biased near the midpoint between the two sweet spots (e.g., 0.3). Thus, similar to the quarton case, when the coupler $|0\rangle\leftrightarrow|1\rangle$ transition is placed above the qubit $|1\rangle\leftrightarrow|2\rangle$ transition, tunable plasmon interactions can be achieved by combining the coupler-mediated coupling with an additional direct qubit coupling (see Appendix~\ref{C}).

Here, we instead consider an alternative scheme that places the qubit's $|1\rangle\leftrightarrow|2\rangle$ transition between the coupler's $|0\rangle\leftrightarrow|1\rangle$ and $|0\rangle\leftrightarrow|2\rangle$ transitions, see Fig.~\ref{fig2}(f). Even without direct qubit coupling, tunable coupling can still be achieved through the following mechanism. (i) As the coupler bias approaches zero flux, the dipole moment of the $|0\rangle\leftrightarrow|2\rangle$ transition gradually vanishes, so the plasmon interaction is dominated by the coupler $|0\rangle\leftrightarrow|1\rangle$ transition and is turned on (see Appendix~\ref{C}). (ii) When the coupler is biased near 0.3, the $|0\rangle\leftrightarrow|1\rangle$ and $|0\rangle\leftrightarrow|2\rangle$  transitions contribute with opposite signs; at a certain bias point they cancel, and the coupling is 
thus turned off (see Appendix~\ref{C}). These observations are confirmed by the results in Figs.~\ref{fig3}(e) and~\ref{fig3}(f). Moreover, as shown in Figs.~\ref{fig3}(g) and~\ref{fig3}(h), both the spectator-induced gate frequency shift and the resulting gate error can also be strongly suppressed when the qubit–spectator coupling is turned off (shaded region).

\section{conclusion}\label{SecV}

To scale fluxonium qubits from few-qubit prototypes to truly large-scale 2D processors, we 
propose an extensible architecture based on low-shunt-capacitance tunable couplers that 
addresses both the capacitance loading issue and spectator errors. This design thus enables 
the coupling infrastructure to be tiled across an entire 2D lattice without sacrificing 
performance. We examine two specific coupler realizations derived from generalized flux-qubit 
circuits. The comparative study could help extract general design principles and motivate further 
investigation of implementation-specific challenges and design refinements. Moreover, the low-shunt-capacitance 
designs could also enable the exploration of more efficient error-correction codes that demand connectivity 
beyond a 2D square lattice~\cite{Breuckmann2021,Bravyi2024,Zhao2026b}.

Looking forward, on the one hand, combining small-footprint qubits such as the fluxonium with equally compact coupler designs may offer a viable route toward miniaturized superconducting quantum processors and denser integration. On the other hand, matching qubit spacing to existing wiring and control technologies could prove challenging; this may be addressed by inserting floating metal islands between qubits and couplers to increase the effective spacing~\cite{Marxer2023,Liang2023}. More importantly, raising the coherence times of non-computational qubit states and couplers from the typical $10\,\mu s$~\cite{Ding2023,Yan2020,Ficheux2021,Liu2023,Azar2026} to several tens of microseconds, and thereby pushing incoherence errors to the $10^{-4}$ level~\cite{Zhao2026,Abad2025}, remains a nontrivial task for such small-footprint circuits~\cite{Gambetta2016}.

\begin{acknowledgments}
We would like to thank Lijing Jin and Guo Xuan Chan for insightful discussions 
on the control wiring fanout issue. The work is supported by the
National Natural Science Foundation of China (Grants No.92576110 and No.12275090). 
Peng Xu is supported by the National Natural Science
Foundation of China (Grants No.12105146 and No.12175104) and the Program of State Key Laboratory of Quantum Optics
Technologies and Devices (No.KF202505).
\end{acknowledgments}

\newpage

\appendix

\section{Full circuit analysis}\label{A}

Here we present a circuit analysis of the studied coupler design, focusing primarily 
on the capacitance loading issue for both floating and grounded fluxonium qubits, as 
well as on spurious capacitive couplings through the circuit itself and their impact 
on coupler functionality. For clarity and without loss of generality, we assume in this work 
that the two island capacitors of the floating qubit are identical and the coupler is of 
a grounded design, see Fig.~\ref{figS1}.

\subsection{Fluxonium-Coupler circuit}\label{A1}

\begin{figure}[tbp]
\begin{center}
\includegraphics[keepaspectratio=true,width=\columnwidth]{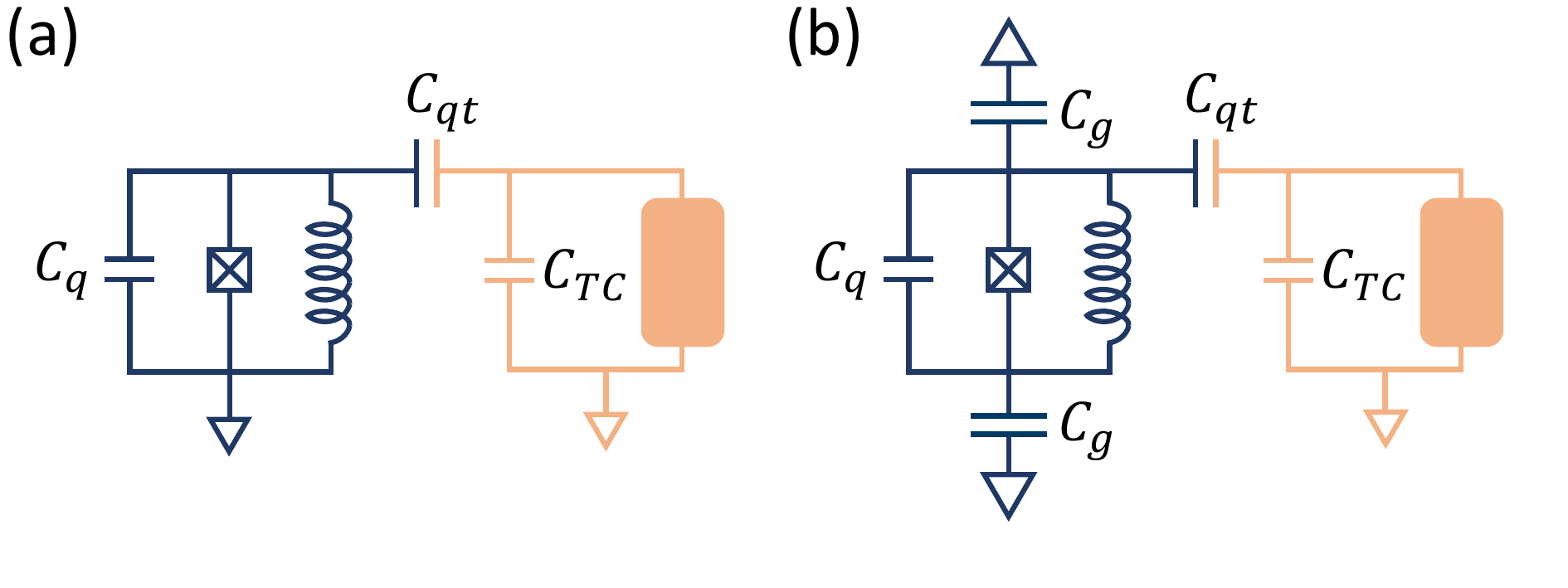}
\end{center}
\caption{(a) Circuit model of a grounded fluxonium qubit (with shunt capacitance $C_{q}$) coupled to a grounded coupler (with shunt capacitance $C_{TC}$) via a coupling capacitor $C_{qt}$. (b) Same as (a) but for a floating fluxonium qubit, with island capacitance $C_{g}$ and shunt capacitance $C_{q}$.}
\label{figS1}
\end{figure}

Given the coupling circuit for the grounded fluxonium qubit shown in Fig.~\ref{figS1}(a), its 
capacitance matrix is
\begin{equation}\label{eqA1}
\mathbf{C_{gn}}=\left(
\begin{array}{cc}
C_{q}+C_{qt} & -C_{qt} \\
-C_{qt}      & C_{TC}+C_{qt}\\
\end{array}
\right),
\end{equation}
while for the floating qubit, see Fig.~\ref{figS1}(b), the capacitance matrix is
\begin{equation}\label{eqA2}
\mathbf{C_{flo}}=\left(
\begin{array}{ccc}
C_g+C_q & -C_q     & 0 \\
-C_q    & C_g+C_q+C_{qt}  & -C_{qt} \\
0 & -C_{qt}        & C_{TC}+C_{qt} \\
\end{array}
\right).
\end{equation}
Note that, in addition to the qubit degree of freedom, the floating qubit circuit 
introduces an additional quantum degree of freedom. This mode does not actually participate in 
the circuit dynamics and can therefore be eliminated by the following transformation of the system variables: 
\begin{equation}\label{eqA3}
\mathbf{S}=\left(
\begin{array}{ccc}
1 & 1 & 0\\
1  & -1& 0 \\
0 & 0 & 1 \\
\end{array}
\right).
\end{equation}
This yields $\mathbf{C_f} = \mathbf{S}^{-1} \mathbf{C_{flo}} \mathbf{S}^{-1}$. After removing the 
free mode, i.e., tracing out the elements relevant to the free mode, we have $\mathbf{[C^{-1}]_{flo}} \equiv \operatorname{Tr}_{\text{F.M.}} \mathbf{C_f}^{-1}$.

The charge energy for each circuit element and their coupling energy can be described 
as follows
\begin{equation}
\begin{aligned}\label{eqA4}
&E_{C,k}\equiv \frac{e^{2}[\mathbf{C}^{-1}]_{k,k}}{2}=\frac{e^{2}}{2C_{\Sigma k}},
\\&J_{jk}\equiv 4e^{2}[\mathbf{C}^{-1}]_{j,k} \,(j\neq k).
\end{aligned}
\end{equation}
Accordingly, the fluxonium–coupler coupling energy is given by
\begin{equation}
\begin{aligned}\label{eqA5}
J_{qt}= \frac{4e^{2}C_{qt}}{C_{TC}C_{q}+C_{qt}(C_{q}+C_{TC})}
\end{aligned}
\end{equation}
for the grounded fluxonium qubit, and by
\begin{equation}
\begin{aligned}\label{eqA6}
J_{qt}=\frac{4e^{2}C_{qt}}{C_{TC}(C_{qt} + \frac{ C_q C_{qt}}{C_{g}} + C_{g}  + 2C_q)+ C_{qt}( C_{g} + 2 C_q)}
\end{aligned}
\end{equation}
for the floating fluxonium qubit. When $C_{qt}\ll \{C_{q},C_{g},C_{TC}\}$, i.e., in the weak-coupling regime, the fluxonium–coupler coupling energy is, to leading order, approximately given by
\begin{equation}
\begin{aligned}\label{eqA7}
J_{qt}\propto\frac{C_{qt}}{C_{TC}},
\end{aligned}
\end{equation}
recovering the formula presented in Eq.~(\ref{eq1}) of the main text.

\subsection{Fluxonium-Coupler-Fluxonium circuit embedded in 2D square lattices}\label{A2}

\begin{table}[!htb]
\caption{\label{tab:Capa_parameters} Capacitor parameters of the coupling circuit for grounded fluxonium qubits embedded in a 2D square lattice similar to that shown in Fig.~\ref{figS2}.}
\begin{ruledtabular}
\begin{tabular}{cccccc}
Capacitance &$C_{TC}$ & $C_{R}$ & $C_{q}$ & $C_{g}$ & $ C_{qr}$ \\\hline
(fF) & 15  & 200 & 7 & 13  & 4 \\
\end{tabular}
\end{ruledtabular}
\end{table}

\begin{table}[!htb]
\caption{\label{tab:Capa_GN_parameters} Capacitor parameters for the Fluxonium–Coupler–Fluxonium circuit embedded in 2D square lattices, as shown in Fig.~\ref{figS2}.}
\begin{ruledtabular}
\begin{tabular}{cccc}
Capacitance & $C_{R}$ & $C_{q}$ & $ C_{qr}$ \\\hline
(fF) & 200 & 13  & 2 \\
\end{tabular}
\end{ruledtabular}
\end{table}

\begin{figure*}[htbp]
\begin{center}
\includegraphics[width=15cm,height=6cm]{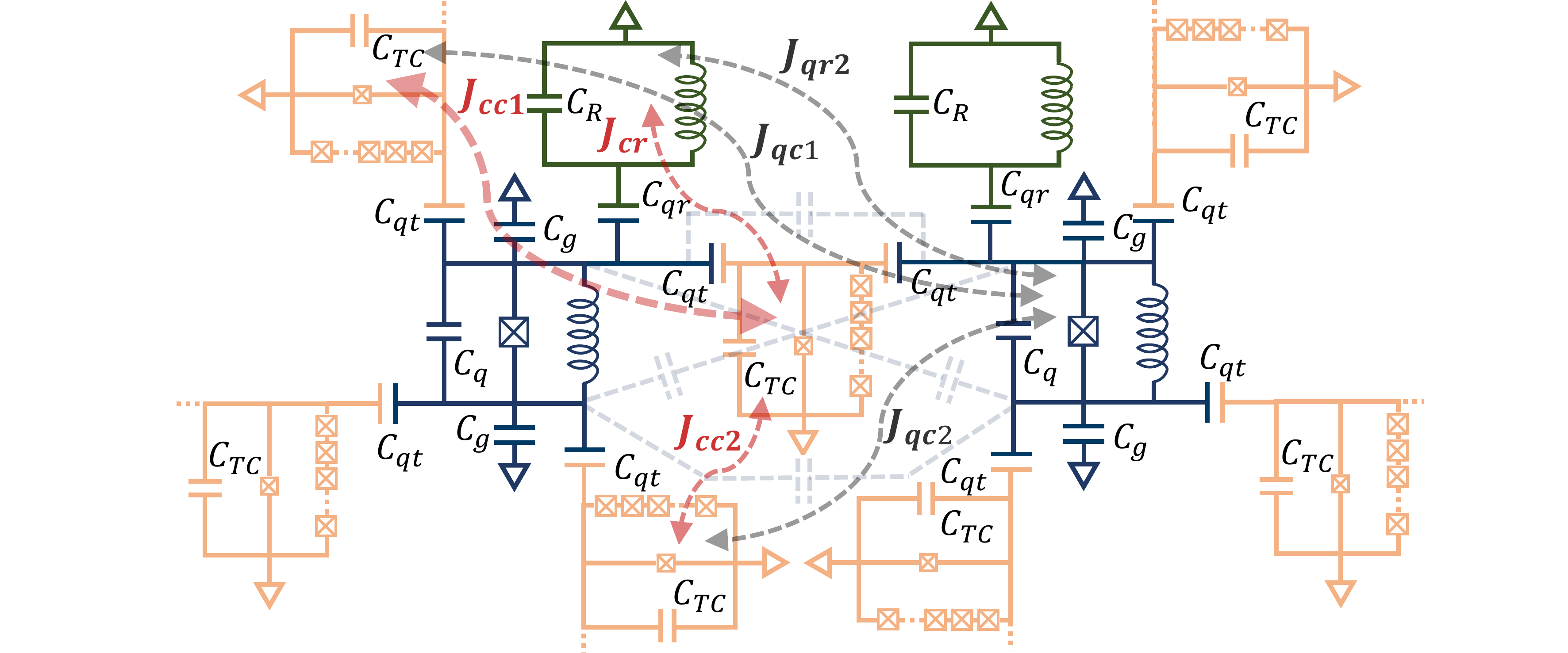}
\end{center}
\caption{Fluxonium-Coupler-Fluxonium circuit embedded in 2D square lattices. For the fluxonium–coupler–fluxonium system, in addition to the fluxonium–coupler couplings, a direct fluxonium–fluxonium coupling (coupling energy $J_{qq}$) also exists, mediated either by the coupling circuit itself or by the capacitance between the fluxonium qubits (dashed blue lines). Double-arrow dashed lines highlight the spurious capacitive couplings among circuit elements (mediated by the coupling circuit itself), including coupler–coupler couplings (two main types, with energies $J_{cc1}$ and $J_{cc2}$, respectively), the coupler–resonator coupling ($J_{cr}$), the coupling between a fluxonium qubit and the readout resonator of a neighboring qubit ($J_{qr2}$, and the coupling between a fluxonium qubit and the coupler of a neighboring qubit (two main types, with energies $J_{qc1}$ and $J_{qc2}$, respectively).}
\label{figS2}
\end{figure*}

\begin{figure}[htbp]
\begin{center}
\includegraphics[keepaspectratio=true,width=\columnwidth]{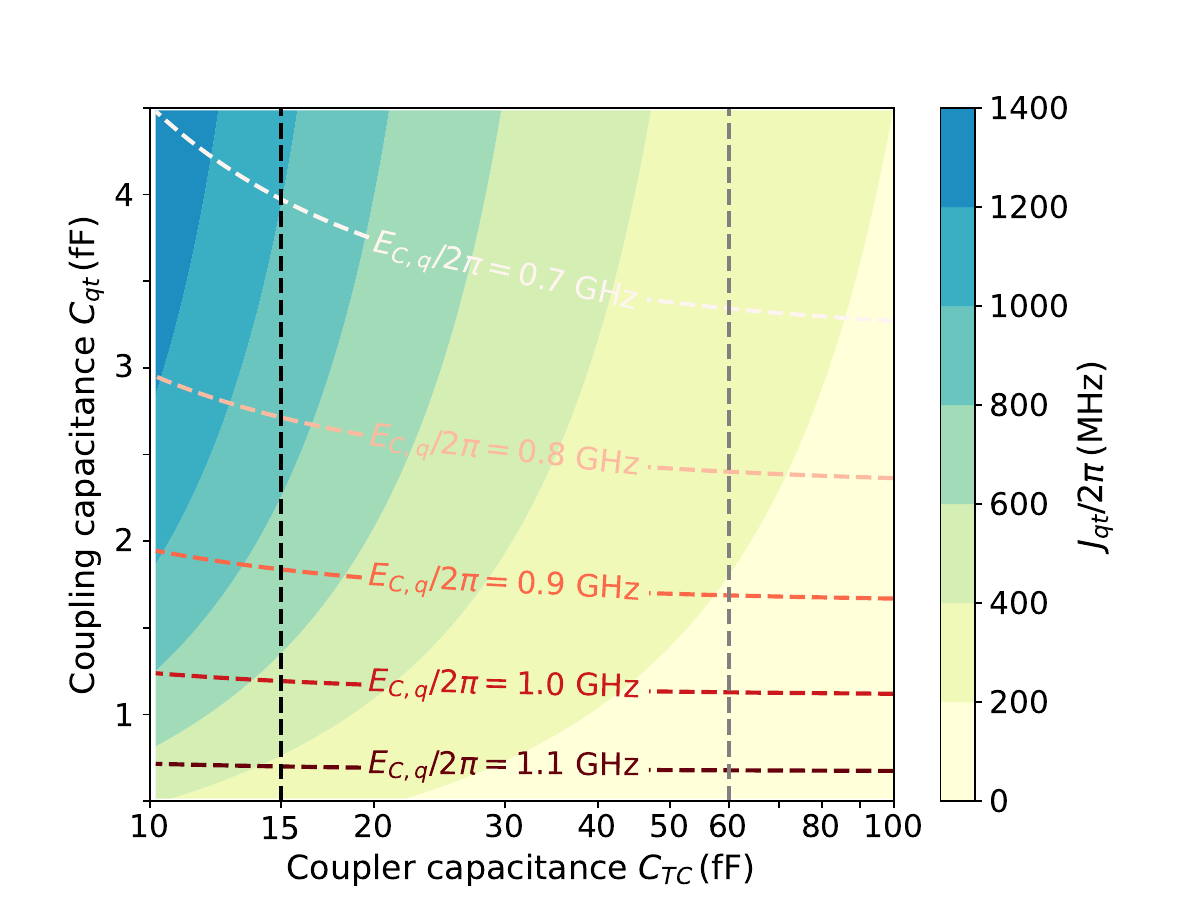}
\end{center}
\caption{Coupling energies $J_{qt}$ (filled contours) and qubit charging energies $E_{C,q}$ (dashed contours) 
as functions of the coupling capacitance $C_{qt}$ and the coupler capacitance $C_{TC}$ for the unit 
cell [as in Fig.~\ref{fig1}(b) of the main text]. The unit cell consists of two grounded fluxonium 
qubits coupled via a grounded coupler, embedded in a 2D square lattice similar to that depicted in Fig.~\ref{figS2}. Vertical dashed grey and black lines mark the typical shunt capacitances for transmon-based coupler designs and for the proposed design, respectively. The capacitor parameters used here are listed in Table~\ref{tab:Capa_GN_parameters}.}
\label{figS3}
\end{figure}

We here examine the capacitance loading issue for floating fluxonium qubits coupled via a grounded 
coupler and focus on a Fluxonium–Coupler–Fluxonium circuit embedded in the 2D square lattice, as 
shown in Fig.~\ref{figS2}. Note that the couplings to the readout resonators and their contribution 
to the qubit charging energies are also taken into account. As in the preceding section, with the 
free mode associated with each floating fluxonium qubit removed, the charging energies of the 
individual circuit elements and the coupling energies among them are given in Eq.~(\ref{eqA4}). 
Accordingly, given the capacitor parameters listed in Table~\ref{tab:Capa_parameters}, we 
obtain both the qubit charging energy and the fluxonium-coupler coupling energy versus 
the coupling capacitance $C_{qt}$ and the coupler shunt capacitance $C_{TC}$ shown in 
Fig.~\ref{fig1}(b) of the main text (see Fig.~\ref{figS3} for the case with grounded fluxonium 
qubits and the capacitor parameters listed in Table~\ref{tab:Capa_GN_parameters}). 

\begin{figure}[htbp]
\begin{center}
\includegraphics[keepaspectratio=true,width=\columnwidth]{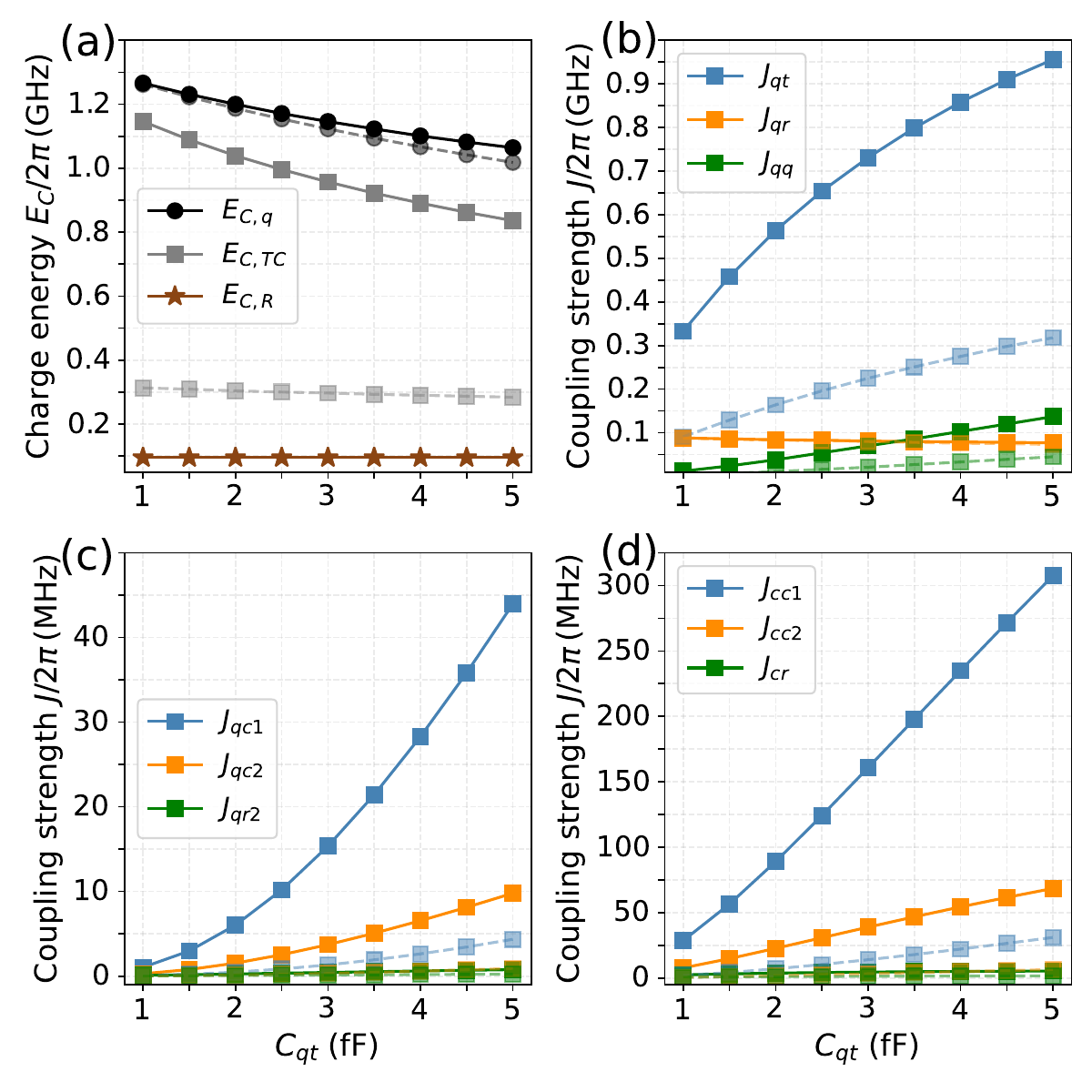}
\end{center}
\caption{(a) Charging energies of the fluxonium qubit, the readout resonator, and the coupler (the one connecting the two fluxonium qubits) versus the coupling capacitance $C_{qt}$. (b) Fluxonium-coupler, direct fluxonium-fluxonium, and fluxonium-resonator coupling energies versus $C_{qt}$. (c) and (d) Spurious coupling energies involving the fluxonium qubit and the coupler, respectively, versus $C_{qt}$. The solid and dashed lines present the results for the proposed coupler design with a shunt capacitance of 15 fF and the transmon-based design with a typical shunt capacitance of 60 fF, respectively.}
\label{figS4}
\end{figure}

In the following, for illustrative purposes, we provide further details on the case of 
floating fluxonium qubits. As shown in Fig.~\ref{figS4}(a), the charging energies of the fluxonium qubit, the coupler (the one between the two fluxonium qubits; see Fig.~\ref{figS2}), and the readout resonator are plotted versus the coupling capacitance $C_{qt}$ for the proposed coupler design with a shunt capacitance of 15 fF. Similarly, the fluxonium–coupler coupling energy $J_{qt}$, the direct fluxonium-fluxonium coupling energy $J_{qq}$, and the fluxonium-resonator coupling energy $J_{qr}$ are also presented, as shown in Fig.~\ref{figS4}(b). For 
easy reference, we also show the results for the transmon-based design with a typical shunt 
capacitance of 60 fF (see lighter dashed lines). Generally, these results indicate that the 
low-shunt-capacitance design can achieve a large coupling energy with a relatively smaller 
coupling capacitance, thereby causing a less severe capacitive loading issue compared to 
designs with a large shunt capacitance.

We note that, besides the target couplings, spurious couplings between circuit elements are also 
present and can potentially compromise the coupler functionality, introducing unwanted quantum 
crosstalk (residual couplings), as discussed below. Figure~\ref{figS2} shows the leading spurious 
couplings (see the double-arrow dashed lines) including the coupler–coupler couplings (two main types, with energies $J_{cc1}$ and $J_{cc2}$, respectively), the coupler–resonator coupling ($J_{cr}$), the coupling 
between a fluxonium qubit and the readout resonator of a neighboring qubit ($J_{qr2}$), and the 
coupling between a fluxonium qubit and the coupler of a neighboring qubit (two main types, with energies $J_{qc1}$ and $J_{qc2}$, respectively). 

Accordingly, Figures~\ref{figS4}(c) and~\ref{figS4}(d) show the magnitudes of these spurious couplings versus the coupling capacitance. Two observations can be made:(i) For coupling circuits that include floating 
elements (here, floating fluxonium qubits), an intrinsic saymmetry exists in the couplings involving 
these floating elements; (ii) In the lattice, the inter-element couplings generally decrease as the 
element–element spacing increases. Intuitively, these can be explained by the fact that these 
spurious couplings are mediated by different capacitor network. For example, the spurious fluxonium-coupler coupling $J_{qc1}$ is mediated by a series of three capacitors ($C_{qt}\xrightarrow{C_{TC}}C_{qt}\xrightarrow{C_{g}}C_{qt}$), whereas $J_{qc2}$ is mediated by a series of four capacitors ($C_{qt}\xrightarrow{C_{TC}}C_{qt}\xrightarrow{C_{g}}C_{q}\xrightarrow{C_{g}}C_{qt}$), hence $J_{qc1}$ is 
generally larger than $J_{qc2}$. A similar consideration applies to the coupler–coupler 
couplings, yielding $J_{cc1}>J_{cc2}$. Bedises, as suggested by Eq.~(\ref{eqA7}), a large shunt capacitance can suppress the coupling energy. For example, owing to the large shunt capacitance of the readout resonator (200 fF; see Table~\ref{tab:Capa_parameters}), the spurious couplings involving the resonator are significantly 
suppressed. 

For ease of comparison, the spurious couplings for the transmon-based design (with a typical shunt capacitance of 60 fF) are also shown (see lighter dashed lines). Specifically, as noted in the main text, achieving a coupling strength of $250\,\rm MHz$ requires $C_{qt}\approx 1.7\,\rm fF$ for the proposed design, whereas the transmon-based design requires $C_{qt}\approx 3.5\, \rm fF$. Accordingly, the spurious coupling energies for the fluxonium qubits are $(J_{qc1},J_{qc2})/2\pi\approx(4,1)\,\rm MHz$ for the proposed design and $(J_{qc1},J_{qc2})/2\pi\approx(2,0.4)\,\rm MHz$ for the transmon-based design. For the coupler, the spurious coupling energies are $(J_{cc1},J_{cc2})/2\pi\approx(69,18)\,\rm MHz$ and $(J_{qc1},J_{qc2})/2\pi\approx(18,4)\,\rm MHz$ for 
the proposed and transmon-based designs, respectively. Given that the transition dipole moment of 
the proposed coupler is typically a factor of two smaller than that of the transmon, we conclude 
that the strengths of these spurious couplings are comparable for the two designs. Moreover, based on 
a rough estimate from Eq.~(\ref{eqA7}), we expect that the strengths of spurious couplings in the popular transmon-based architecture (transmon qubits are coupled via a transmon-based tunable 
coupler~\cite{Yan2018}) would also be of a comparable level.

\subsection{Spurious coupler-coupler couplings}\label{A3}

\begin{figure}[tbp]
\begin{center}
\includegraphics[keepaspectratio=true,width=\columnwidth]{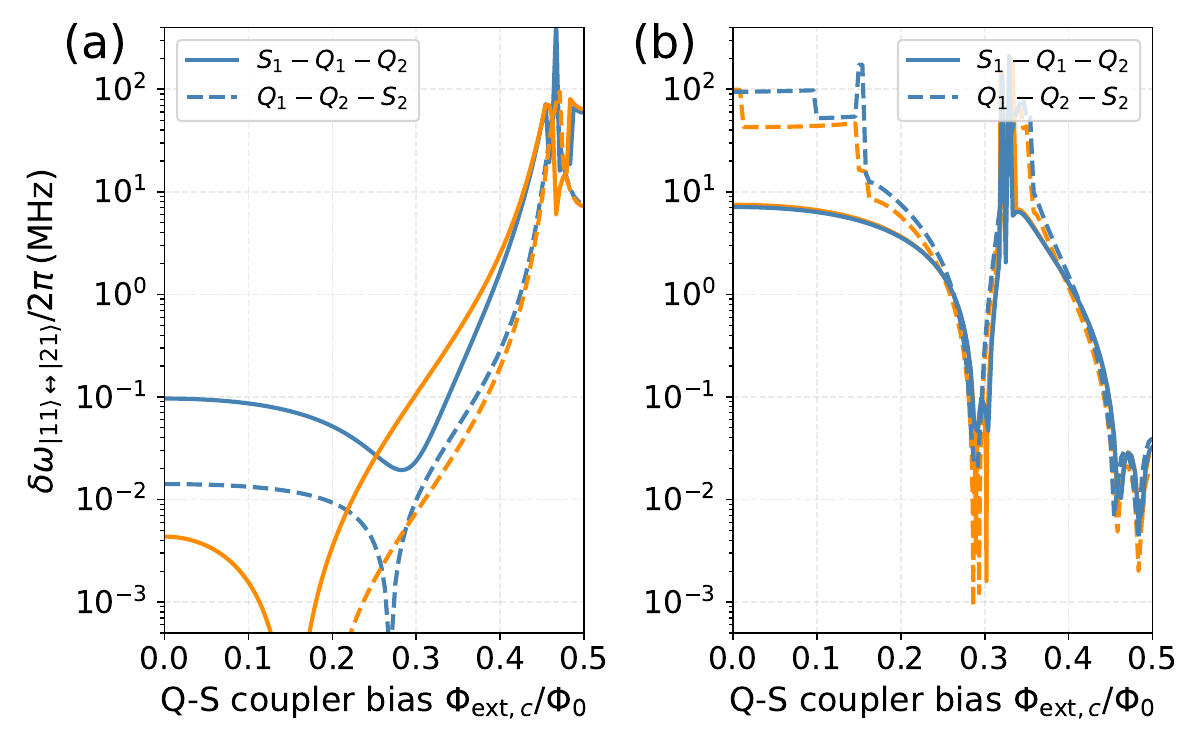}
\end{center}
\caption{Frequency shift of the gate transition induced by residual spectator-qubit coupling, shown as a function of the flux bias applied to the qubit-spectator coupler for (a) the quarton-circuit design and (b) the fluxonium-circuit design [as in Figs.~\ref{fig3}(c) and~\ref{fig3}(g) of the main text]. Solid and dashed lines correspond to the configurations $S_{1}-Q_{1}-Q_{2}$ and $Q_{1}-Q_{2}-S_{2}$, respectively. Blue and orange lines represent the results with and without (corresponding to the results already presented in Figs.~\ref{fig3}(c) and~\ref{fig3}(g) of the main text) a spurious coupler–coupler coupling of $100\,\rm MHz$.}
\label{figS5}
\end{figure}

From Figs.~\ref{figS4}(c) and~\ref{figS4}(d), we conclude that the dominant spurious couplings are the coupler–coupler couplings, whose strength can reach $\sim 100\,\rm MHz$ when the fluxonium–coupler coupling energy is targeted at $\sim 550\,\rm MHz$. To examine their impact on the coupler functionality, we present in Figs.~\ref{figS5}(a) and~\ref{figS5}(b) the frequency shift of the gate transition induced by spectator-qubit coupling, now including a coupler–coupler coupling energy of $100\,\rm MHz$ [similar to the analysis shown in Figs.~\ref{fig3}(c) and~\ref{fig3}(g) of the main text]. For ease of reference, the case without the coupler–coupler coupling is also shown (orange lines), corresponding to the results already presented in Figs.~\ref{fig3}(c) and~\ref{fig3}(g) of the main text. We find that, in general, the inclusion of such spurious couplings does not fundamentally alter the coupler functionality (i.e., the tunable control of plasmon interactions) but does affect the coupling-off bias points. Moreover, comparing the two coupler designs, the impact of spurious couplings on the fluxonium-circuit coupler is significantly weaker than on the quarton-circuit coupler. This can be partly explained by the fact that the transition dipole moment of the fluxonium-circuit coupler at the coupling-off bias point (0.429; see Table~\ref{tab:transition_mag_parameters}) is smaller than that of the quarton coupler (0.870; see the same table), thereby yielding weaker spurious coupling strengths.

Finally, we note that the spurious coupling issue is not unique to the fluxonium architecture, but 
is similarly encountered in widely used transmon architectures with tunable couplers (see, e.g., Ref.~\cite{Zajac2021}). This suggests that, even with tunable coupling, achieving high system 
performance requires engineering both the couplings between neighboring qubits and the 
interactions among all potentially coupled spectator qubits or couplers~\cite{Klimov2024}, regardless of whether 
the qubit architecture comprises fluxonium or transmon qubits.

\section{Coupler-induced decoherence}\label{B}

\begin{figure}[htbp]
\begin{center}
\includegraphics[keepaspectratio=true,width=\columnwidth]{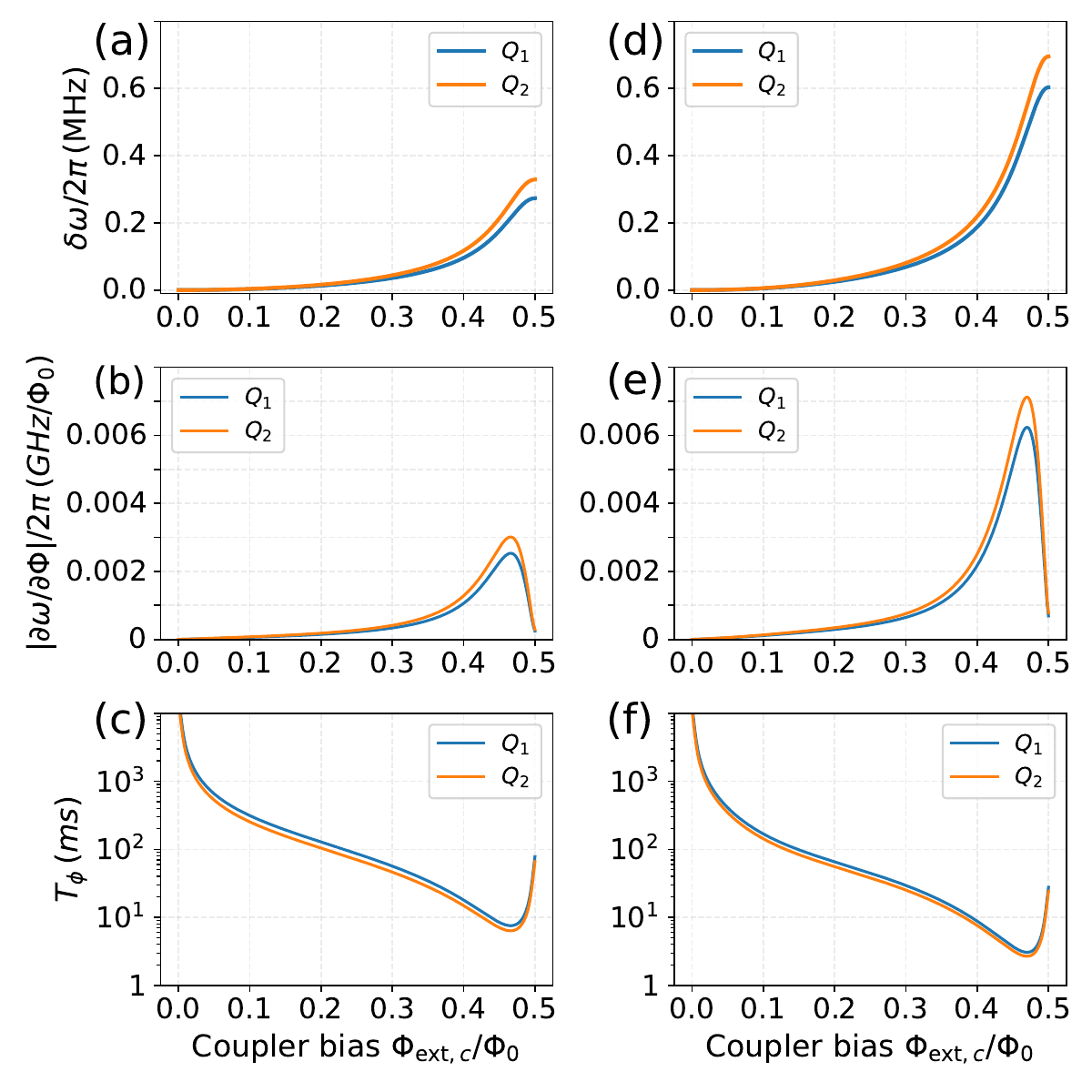}
\end{center}
\caption{(a,d) The coupler-induced frequency shift in the fluxonium's qubit transition ($|0\rangle\leftrightarrow|1\rangle$), (b,e) the sensitivity of the qubit transition frequency to the coupler flux bias $\partial\omega/\partial\Phi$, and (c,f) the coupler-induced pure dephasing as functions of the coupler flux bias. (a-c) and (d-f) are for the quarton-circuit coupler and the fluxonium-circuit coupler, respectively. Here, the dephasing times are estimated by assuming 1/f noise with the noise amplitude of $10\,\mu\Phi_{0}$. }
\label{figS6}
\end{figure}

\begin{figure}[htbp]
\begin{center}
\includegraphics[keepaspectratio=true,width=\columnwidth]{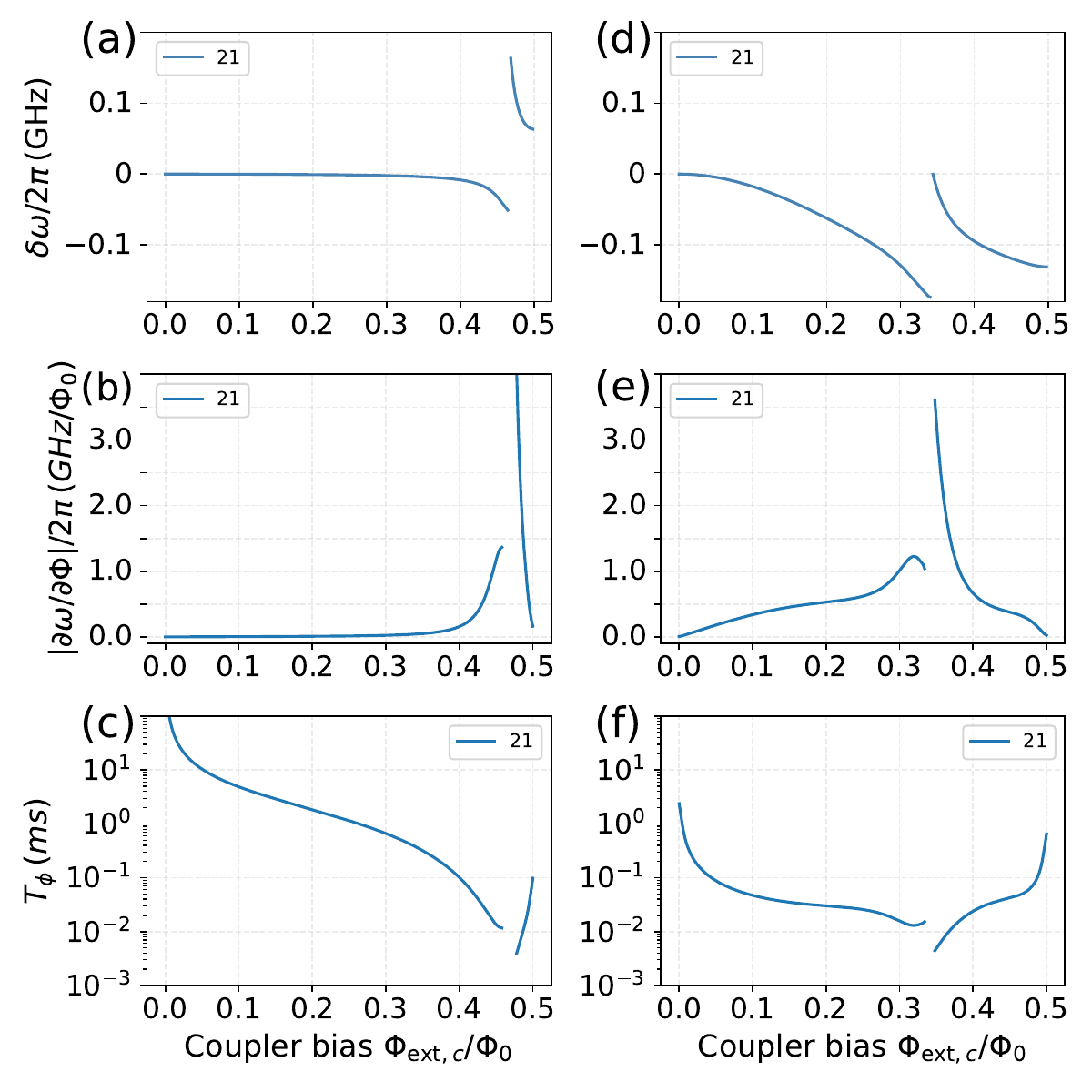}
\end{center}
\caption{(a,d) The coupler-induced frequency shift in the non-computational gate state $|21\rangle$, (b,e) the sensitivity of the energy of the non-computational gate state to the coupler flux bias $\partial\omega/\partial\Phi$, and (c,f) the coupler-induced pure dephasing as functions of the coupler flux bias. (a-c) and (d-f) are for the quarton-circuit coupler and the fluxonium-circuit coupler, respectively. Here, the dephasing times are estimated by assuming 1/f noise with the noise amplitude of $10\,\mu\Phi_{0}$. Note that discontinuities in curves are caused by state labeling failure near avoided crossings.}
\label{figS7}
\end{figure}

\begin{figure}[htbp]
\begin{center}
\includegraphics[keepaspectratio=true,width=\columnwidth]{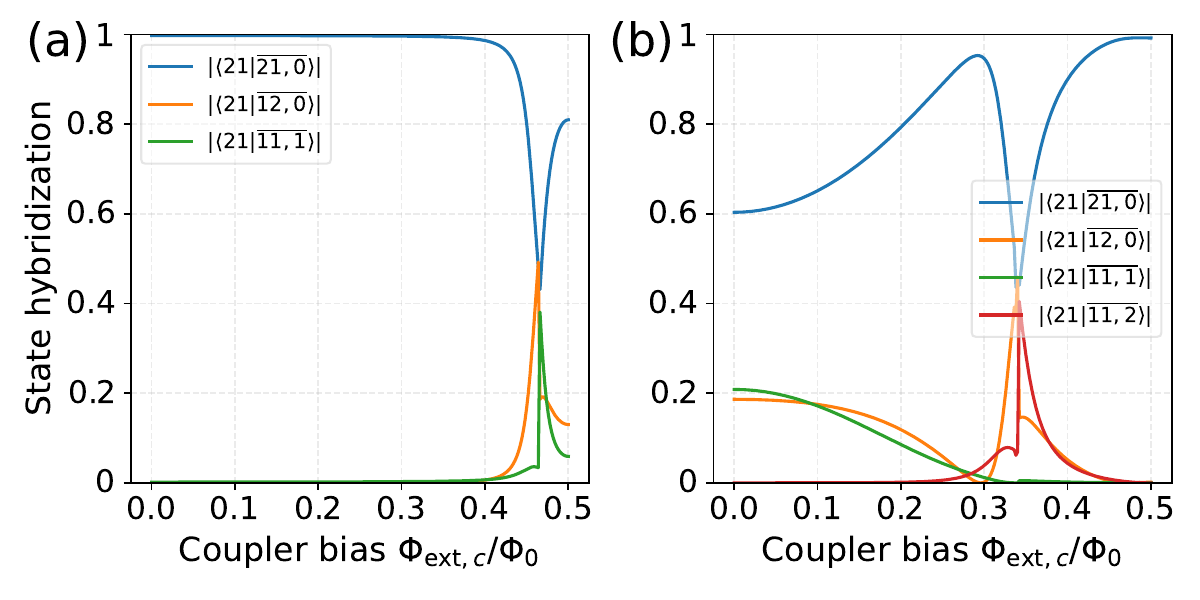}
\end{center}
\caption{State hybridization of the non-computational gate state $|21\rangle$ induced by its coupling to the coupler, quantified by the overlap between the state and the corresponding bare states. Panels (a) correspond to the quarton-circuit design, and (b) to the fluxonium-circuit design. Sudden jumps or discontinuities in the curves are caused by state labeling failure near avoided crossings.}
\label{figS8}
\end{figure}

Following the same analysis framework as in Ref.~\cite{Zhao2026}, we now examine the 
coupler-induced dephasing and relaxation for the proposed coupling architecture. Given 
the general applicability of that analysis, we focus primarily on summarizing the key 
findings, both to avoid repetition and to highlight any architecture-specific 
considerations.

We first focus on the computational subspace. Figure~\ref{figS6} shows the coupler-induced 
frequency shift of the qubit transition and the resulting qubit dephasing for both coupling 
architectures, i.e., the one with the quarton-circuit coupler and that with the fluxonium-circuit 
coupler. As noted in the main text, owing to the weak transition dipole moment of the qubit transition and the low transition frequency, the qubit subspace is almost completely decoupled from the coupler. This effective decoupling strongly suppresses the Purcell-like decay through the coupler and the coupler-induced dephasing (dephasing 
times are estimated assuming 1/f noise with the noise amplitude of $10\,\mu\Phi_{0}$~\cite{Nguyen2019}, see Figs.~\ref{figS6}(c) and.~\ref{figS6}(f) for the two coupler designs, respectively). Consequently, similar to the transmon-based architecture proposed in Ref.~\cite{Zhao2026}, the coupling architecture studied here should also maintain the state-of-the-art coherence already demonstrated in the computational subspace of fluxonium qubits.

In the coupling architecture studied here, the two-qubit gate is realized by activating transitions to non-computational states following the MAP gate scheme. As a result, the non-computational states of the fluxonium qubit are temporarily populated during gate operations. The coupler-induced decoherence on these states must also be examined, as in Ref.~\cite{Zhao2026}. In the present work, we focus on the gate transition  $|11\rangle\leftrightarrow|21\rangle$, and, for illustrative purposes, we thus evaluate the coupler-induced 
decoherence for the non-computational state $|21\rangle$ below. For both coupler designs, Figure~\ref{figS7} shows the coupler-induced frequency shift of the non-computational state $|21\rangle$ and the resulting dephasing, while Figure~\ref{figS8} presents the state hybridization of $|21\rangle$ induced by its coupling to the coupler, from which the coupler-induced Purcell-like decay can be inferred. 

As shown in Fig.~\ref{figS7}(c), for the quarton-circuit coupler, which is biased near 0.5 (specifically, at 0.456, see Table~\ref{tab:QC_bias_parameters}) to turn on the plasmon interaction (i.e., the coupler interaction point) during gate operations, the coupler-induced dephasing time ($T_{\phi,{\rm 1/f}}$) is generally above $5\,\rm \mu s$ (assuming $1/f$ noise with a typical noise amplitude of $10\,\mu\Phi_{0}$) at this bias point. For 50-ns CZ gates, this dephasing leads to an incoherent error of $\epsilon\approx (13/80)(t_g/T_{\phi,{\rm 1/f}})^2\approx1.6\times10^{-5}$~\cite{Zhao2026} and can thus be safely ignored. In contrast, for the fluxonium-circuit coupler, the coupler-induced dephasing time is generally 
above $50\,\rm \mu s$ [see Fig.~\ref{figS7}(f)] at the coupler interaction point (near the zero-bias point), and the resulting incoherent error can also be neglected. 

Unlike the coupler-induced dephasing, the gate error arising from the coupler-induced Purcell-like decay could be non-negligible, as illustrated below. As shown in Fig.~\ref{figS8}, at the respective coupler interaction point, the maximum state overlap reaches approximately $40\%$ for the quarton-circuit coupler and $20\%$ for the 
quarton-circuit coupler. We note that the typical coupler relaxation times $T_{1,c}$ are currently $10-50\,\rm \mu s$~\cite{Yan2020} for the quarton circuit (biased at the half-integer flux quantum) and is $10-30\,\rm \mu s$~\cite{Liu2023,Azar2026} for the fluxonium circuit (around the zero-bias point), while the reported lifetime 
of the non-computational gate state is on the order of $10\,\rm \mu s$~\cite{Ficheux2021,Ding2023}. 
For 50-ns CZ gates, assuming a worst-case scenario (state overlap of $40\%$ and coupler relaxation time $10\,\rm \mu s$), this Purcell-like decay yields an incoherent error $\epsilon\approx (0.4)^{2}(3/32)(t_g/T_{1,c})\approx7.5\times10^{-5}$~\cite{Zhao2026}. Thus, we conclude that Purcell decay does not currently appear to limit the gate performance.

\section{Effective tunable coupling}\label{C}

Here, we present an effective model that captures the tunable control of the plasmon interaction $|11\rangle\leftrightarrow|22\rangle$ for both tunable coupler designs. We consider a system composed of two fluxonium qubits, $Q_{1}$ and $Q_{2}$, coupled via a tunable coupler (either a quarton-circuit or a fluxonium-circuit coupler). The full system Hamiltonian is given by (i.e., Eq.~(\ref{eq4}) of the main text)
\begin{equation}
\begin{aligned}\label{eqC1}
\hat{H}= &\sum_{j=1,2} [\hat{H}_{fj}+J_{cj} \hat{n}_{fj}\hat{n}_{c}]+\hat{H}_{c}+J_{12}\hat{n}_{f1} \hat{n}_{f2}.
\end{aligned}
\end{equation}
Since we focus on the plasmon transition $|1\rangle\leftrightarrow|2\rangle$, we define the corresponding lowering and raising operators for qubit $Q_{j}$ as
\begin{equation}
\begin{aligned}\label{eqC2}
&\hat p_{j}=[|1\rangle\langle 2|]_{j},\,\hat p_{j}^{\dag}=[|2\rangle\langle 1|]_{j}.
\end{aligned}
\end{equation}

\subsection{Quarton-circuit coupler}\label{C1}

As mentioned in the main text, for the quarton-circuit coupler we focus on its lowest transition $|0\rangle\leftrightarrow|1\rangle$, while higher-level transitions are omitted because their 
frequencies lie far beyond the range relevant to our study. Accordingly, we define the 
corresponding lowering and raising operators for the coupler as
\begin{equation}
\begin{aligned}\label{eqC3}
&\hat c_{q}=[|0\rangle\langle 1|]_{j},\,\hat c_{q}^{\dag}=[|1\rangle\langle 0|]_{j}.
\end{aligned}
\end{equation}

Given the system Hamiltonian in Eq.~(\ref{eqC1}) and focusing on the fluxonium transition $|1\rangle\leftrightarrow|2\rangle$, the Hamiltonian can be rewritten as
\begin{equation}
\begin{aligned}\label{eqC4}
\hat H_{p}=&\sum_{j=1,2}\left[\omega_{p,j}\hat p_{j}^{\dag}\hat p_{j}+g_{p,j}(\hat p_{j}+\hat p_{j}^{\dag})(\hat c_{q}+\hat c_{q}^{\dag})\right]
\\&+\omega_{c,q}\hat c_{q}^{\dag}\hat c_{q}+g_{p,12}(\hat p_{1}+\hat p_{1}^{\dag})(\hat p_{2}+\hat p_{2}^{\dag}),
\end{aligned}
\end{equation}
where
\begin{equation}
\begin{aligned}\label{eqC5}
&g_{p,j}=J_{cj}|\langle 2|\hat n_{fj}| 1\rangle \langle 1 |\hat n_{c}|0\rangle|,
\\&g_{p,12}=J_{12}|\langle 2 |\hat n_{f1}|1\rangle \langle 2|\hat n_{f2}|1\rangle|,
\end{aligned}
\end{equation}
are the plasmon–coupler and direct plasmon–plasmon coupling strengths, respectively. Here, $\omega_{p,j}$ and $\omega_{c,q}$ denote the fluxonium plasmon transition frequency and the coupler transition 
frequency, respectively.

Considering that the coupled plasmon–coupler system operates in the dispersive regime, i.e., when $g_{p,j}$ is much smaller than the plasmon–coupler detuning $\Delta_{p,j}=|\omega_{p,j}-\omega_{c,q}|$, an effective interaction Hamiltonian can be derived by eliminating the direct plasmon–coupler interactions~\cite{Yan2018,Niskanen2007,Mundada2019}. This yields
\begin{equation}
\begin{aligned}\label{eqC6}
\hat H_{p,{\rm eff}}^{(I)}= g_{p}(\hat p_{1}\hat p_{2}^{\dag}+\hat p_{1}^{\dag}\hat p_{2}),
\end{aligned}
\end{equation}
which describes the coupler-mediated plasmon–plasmon interaction. The corresponding coupling strength is given by
\begin{equation}
\begin{aligned}\label{eqC7}
g_{p}=g_{p,12}+\frac{g_{p,1}g_{p,2}}{2}\left(\frac{1}{\Delta_{p,1}}+\frac{1}{\Delta_{p,2}}\right).
\end{aligned}
\end{equation}

\begin{figure*}[htbp]
\begin{center}
\includegraphics[width=15cm,height=8cm]{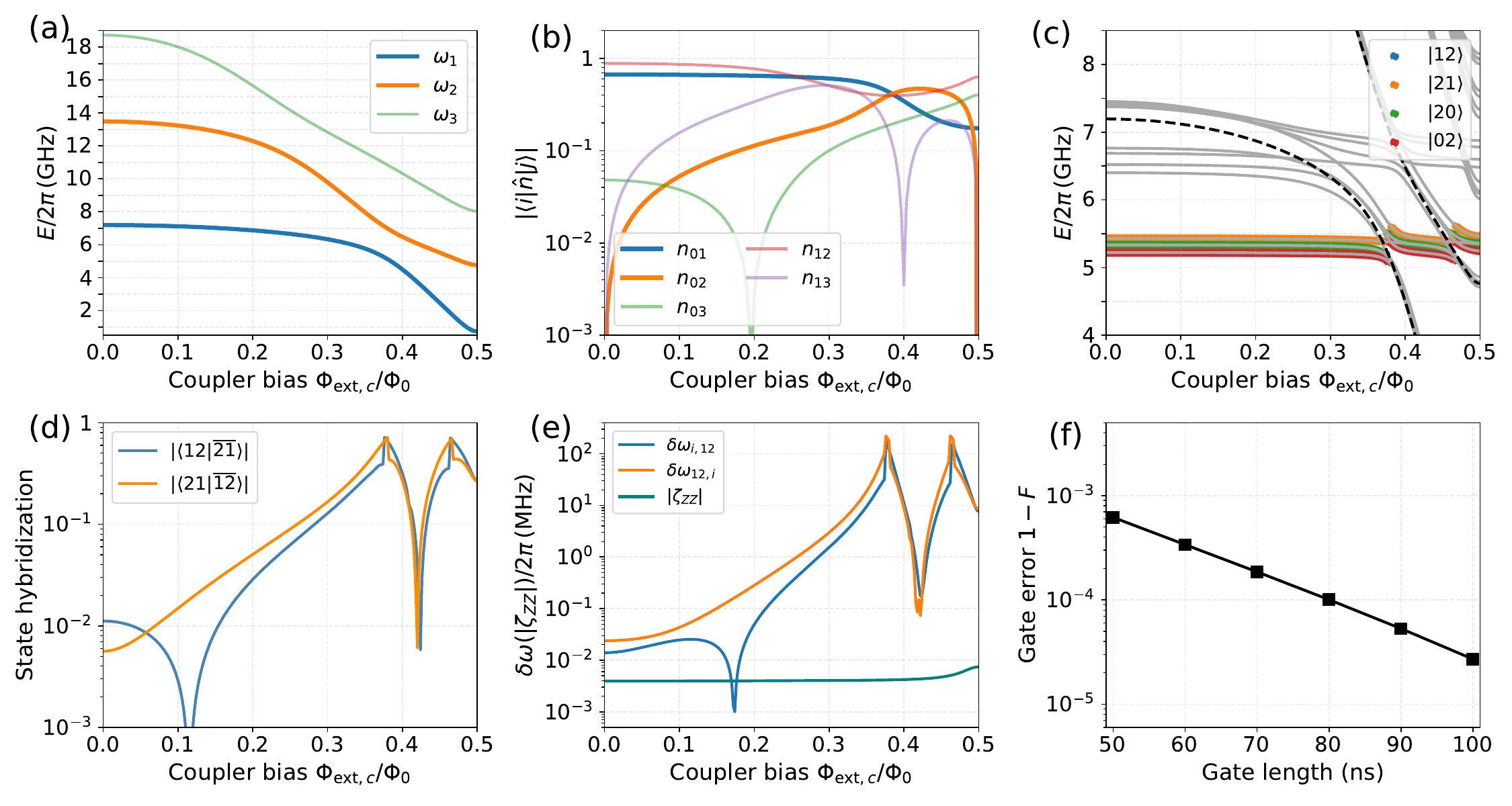}
\end{center}
\caption{ Analysis of the fluxonium-circuit coupler design (similar to the quarton-circuit coupler) with the parameters listed in Table~\ref{tab:FC2_parameters}. (a) Energy levels of the fluxonium-circuit coupler, (b) magnitudes of the electric transition dipole moments for the coupler transitions, and (c) full energy levels of the coupled system (two fluxonium qubits coupled via the coupler) as functions of the coupler flux bias. The non-computational energy levels $\{|20\rangle,|02\rangle,|21\rangle,|12\rangle\}$ associated with the plasmon transition $|1\rangle \leftrightarrow |2\rangle$ are highlighted, together with the bare coupler frequency (black dashed lines, as in (a)). (d) State hybridization and (e) state-dependent frequency shift of the transition $|1\rangle\leftrightarrow|2\rangle$ versus coupler bias. (f) CZ gate errors as a function of gate length (here, the coupler is biased 
at 0.373, leading to a state-dependent frequency shift of $92.307\,\rm MHz$). In (d) and (e), abrupt jumps originate from state-labeling failures near avoided crossings.}
\label{figS9}
\end{figure*}

\begin{table}[!htb]
\caption{\label{tab:FC2_parameters} Parameters of the fluxonium-circuit coupler used in the numerical analysis shown in Fig.~\ref{figS9}.}
\begin{ruledtabular}
\begin{tabular}{cccc}
(GHz) & $E_C/2\pi$& $E_L/2\pi$& $E_J/2\pi$ \\\hline
Fluxonium coupler & 1.0 & 2.0 & 6.0\\
\hline
\hline
(MHz) & $J_{c1}/2\pi$  & $J_{c2}/2\pi$  & $J_{12}/2\pi$ \\\hline
Coupling strengths  & 650 & 650 & 110
\end{tabular}
\end{ruledtabular}
\end{table}

As indicated by the above analysis, the flip-flop interactions between plasmon transitions can be turned off or on by tuning the quarton transition frequency $\omega_{c,q}$ (i.e., the plasmon-coupler 
detuning $\Delta_{p,j}$). Similarly, a tunable coupler design that combines the coupler-mediated coupling with an additional direct qubit coupling can also be realized using the fluxonium-circuit coupler. Specifically, Figure~\ref{figS9} shows the numerical results for this design, obtained with the parameters listed in Table~\ref{tab:FC2_parameters}. Note that unlike the tunable control of computational-state interactions demonstrated in Refs.~\cite{Moskalenko2021,Moskalenko2022}, which is limited by the weak transition dipole in the qubit subspace, we instead focus here on interactions involving non-computational states.

\subsection{Fluxonium-circuit coupler}\label{C2}

For the fluxonium-circuit coupler discussed in the main text, we focus on its lowest two transitions, $|0\rangle\leftrightarrow|1\rangle$ and $|0\rangle\leftrightarrow|2\rangle$, while higher transitions are omitted because their frequencies lie far beyond the range relevant to our study. Accordingly, we define the corresponding lowering and raising operators for these two transitions as
\begin{equation}
\begin{aligned}\label{eqC8}
&\hat c_{q1}=[|0\rangle\langle 1|]_{j},\,\hat c_{q1}^{\dag}=[|1\rangle\langle 0|]_{j},
\\&\hat c_{q2}=[|0\rangle\langle 2|]_{j},\,\hat c_{q2}^{\dag}=[|2\rangle\langle 0|]_{j}.
\end{aligned}
\end{equation}
Moreover, as mentioned in the main text, this tunable coupler design does not require a direct fluxonium–fluxonium coupling; in the following discussion, we therefore neglect the direct coupling terms in Eq.~\eqref{eqC1}.

As in the preceding analysis, when we focus on the plasmon transition $|1\rangle\leftrightarrow|2\rangle$,  the system Hamiltonian can be rewritten as
\begin{equation}
\begin{aligned}\label{eqC9}
\hat H_{p}=&\sum_{j=1,2}\left[\omega_{p,j}\hat p_{j}^{\dag}\hat p_{j}\right]+\sum_{k=1,2}\left[\omega_{c,qk}\hat c_{qk}^{\dag}\hat c_{qk}\right]
\\&+\sum_{j,k=1,2}\left[g_{p,jk}(\hat p_{j}+\hat p_{j}^{\dag})(\hat c_{qk}+\hat c_{qk}^{\dag})\right],
\end{aligned}
\end{equation}
where
\begin{equation}
\begin{aligned}\label{eqC10}
&g_{p,jk}=J_{cj}|\langle 2|\hat n_{fj}| 1\rangle \langle k |\hat n_{c}|0\rangle|
\end{aligned}
\end{equation}
denotes the strength of the coupling between the plasmon transition $|1\rangle\leftrightarrow|2\rangle$ and the coupler transition $|0\rangle\leftrightarrow|k\rangle$ (with transition frequency $\omega_{c,qk}$).

Again, when the coupled plasmon–coupler system operates in the dispersive regime, i.e., when $g_{p,jk}$ is much smaller than the plasmon–coupler detuning $\Delta_{p,jk}=|\omega_{p,j}-\omega_{c,qk}|$, an effective interaction Hamiltonian can be obtained by eliminating the direct plasmon–coupler interactions, leading to
\begin{equation}
\begin{aligned}\label{eqC11}
\hat H_{p,{\rm eff}}^{(I)}= g_{p}(\hat p_{1}\hat p_{2}^{\dag}+\hat p_{1}^{\dag}\hat p_{2}),
\end{aligned}
\end{equation}
which describes the coupler-mediated plasmon–plasmon interaction. The corresponding coupling strength is given by
\begin{equation}
\begin{aligned}\label{eqC12}
g_{p}=\sum_{k=1,2}\left[\frac{g_{p,1k}g_{p,2k}}{2}\left(\frac{1}{\Delta_{p,1k}}+\frac{1}{\Delta_{p,2k}}\right)\right],
\end{aligned}
\end{equation}
where the two terms in the sum correspond to the mediated plasmon interactions arising from the coupler's two transitions.

When the plasmon transition $|1\rangle\leftrightarrow|2\rangle$ is placed between the coupler's $|0\rangle\leftrightarrow|1\rangle$ and $|0\rangle\leftrightarrow|2\rangle$ transitions, we 
have $\Delta_{p,j1}>0$ and $\Delta_{p,j2}<0$. Thus, according to Eq.~(\ref{eqC12}), the two terms contribute to the effective plasmon interaction with opposite signs and, at a certain bias point, they cancel each other, thereby turning off the plasmon interaction. In contrast, when the coupler is biased to the zero-bias point, the contribution from the $|0\rangle\leftrightarrow|2\rangle$ transition is suppressed because this transition is forbidden at that point, while the contribution from the $|0\rangle\leftrightarrow|1\rangle$ transition is maintained, thus turning on the plasmon interaction.

\section{MAP-based CZ gates}\label{D}

\begin{table}[!htb]
\caption{\label{tab:qubit_fre_parameters} Bare fluxonium and coupler transition frequencies based on the circuit parameters listed in Table~\ref{tab:circuit_parameters}. For the coupler, values without parentheses refer to the idle point, while values in parentheses refer to the interaction point.}
\begin{ruledtabular}
\begin{tabular}{cccc}
$ $&
$\omega^{(01)}/2\pi$ (GHz)&
$\omega^{(12)}/2\pi$ (GHz)&
$\omega^{(03)}/2\pi$ (GHz)\\\hline
$Q_{1}$ & 0.114 & 5.253 & 6.802 \\
$Q_{2}$ & 0.097 & 5.167 & 6.501  \\
$S_{1}$ & 0.096 & 5.185 & 6.529 \\
$S_{2}$ & 0.113 & 5.281 & 6.827 \\
$C_{Q}$  & 12.123 (5.618) & 11.585 (6.037)  & 34.755 (18.536) \\
$C_{F}$ & 3.806 (4.826) & 1.976 (3.847) & 8.676 (11.451) 
\end{tabular}
\end{ruledtabular}
\end{table}

\begin{table}[!htb]
\caption{\label{tab:transition_mag_parameters} Magnitudes of the transition matrix elements for the plasmon transition and the coupler, based on the circuit parameters listed in Table~\ref{tab:circuit_parameters}. For the coupler, values without parentheses refer to the idle point, while values in parentheses refer to the interaction point.}
\begin{ruledtabular}
\begin{tabular}{ccccc}
$ $& $n_{01}$ & $n_{12}$ & $n_{02}$ & $n_{03}$ \\\hline
$Q_{1}$ & 0.038 & 0.603 & 0 & 0.582 \\
$Q_{2}$ & 0.033 & 0.595 & 0 & 0.576  \\
$S_{1}$ & 0.033 & 0.596 & 0 & 0.578 \\
$S_{2}$ & 0.038 & 0.604 & 0 & 0.584 \\
$C_{Q}$  & 0.870 (0.569) & 1.202 (0.843)  & 0 (0.218) & 0.023 (0.114)\\
$C_{F}$ & 0.429 (0.548) & 0.437 (0.686) & 0.272 (0) & 0.105 (0.060)
\end{tabular}
\end{ruledtabular}
\end{table}

Here, we provide a detailed description of the microwave-activated CZ gate based on 
the MAP scheme~\cite{Chow2013}, together with the procedure for characterizing intrinsic gate errors 
and spectator-induced gate errors. As in Ref.~\cite{Zhao2026}, tunable plasmon interactions can be leveraged to engineer state-dependent frequency shifts, thereby realizing microwave-activated CZ gates~\cite{Nesterov2018,Ficheux2021}. Specifically, the plasmon-mediated interaction can enable selective transitions, such as the $|21\rangle\leftrightarrow|12\rangle$ studied here. Following the MAP scheme, the qubit state (here $|11\rangle$) accumulates an additional phase of $\pi$ when the system undergoes a full Rabi oscillation~\cite{Chow2013,Nesterov2018}. Thus, a CZ gate is realized after correcting the accumulated single-qubit phases, typically via virtual-Z gates with near-perfect fidelity and zero duration~\cite{McKay2017}.

In the proposed coupling architecture, a CZ gate can therefore be implemented by first tuning the coupler 
from its idle point (turning off the plasmon interaction) to the interaction point (turning on the plasmon interaction), then waiting for one period of the selectively driven Rabi oscillation, and finally biasing the coupler back to the idle point~\cite{Zhao2026}. We note that, due to the near-complete decoupling between the computational subspace and the coupler, non-adiabatic transitions of computational states during the coupler bias ramping (i.e., from idle to interaction point, and vice versa) can be neglected. For clarity, we neglect the bias ramping process in 
the following analysis, while noting that in actual implementations such ramping inevitably takes a 
nonzero duration (to accommodate control electronics constraints in practical systems)~\cite{Zhan2026}. 
For easy reference, based on the system Hamiltonian parameters listed in Table~\ref{tab:circuit_parameters}, we summarize the bare fluxonium transition frequencies and the coupler transition frequency (at both the idle and the interaction point) in Table~\ref{tab:qubit_fre_parameters}, and the magnitudes of the transition matrix elements for the plasmon modes and the coupler in Table~\ref{tab:transition_mag_parameters}.

Following Refs.~\cite{Nesterov2018,Ding2023}, we consider identical microwave drives (same amplitude and frequency) applied simultaneously to two coupled fluxonium qubits to activate the gate transition (here $|11\rangle\leftrightarrow|21\rangle$). The driven Hamiltonian is written as
\begin{equation}
\begin{aligned}\label{eqD1}
H_{d}=\sum_{i=1,2}A(t)\cos(\omega_{d}t+\phi_{i})\hat n_{i},
\end{aligned}
\end{equation}
where $A$, $\omega_{d}$, and $\phi_{i}$ denotes the amplitude, frequency, and the
phase of the drive, respectively. The relative phase between the two drives is determined by minimizing the period of the activated Rabi oscillation~\cite{Ding2023}. To mitigate spurious transitions, we adopt the derivative removal by adiabatic gate (DRAG) scheme~\cite{Motzoi2009}. The full DRAG pulse is then given by
\begin{equation}
\begin{aligned}\label{eqD2}
A_{DRAG}(t)=A(t)+i\frac{\alpha}{\delta\omega_{12,i}} \frac{dA(t)}{dt},
\end{aligned}
\end{equation}
with $\alpha=1$ to minimize the leading spurious transition (here $|10\rangle\leftrightarrow|20\rangle$), which is detuned from the target gate transition by $\delta\omega_{12,i}$. Here, we use a cosine-shaped pulse, i.e.,
\begin{equation}
\begin{aligned}\label{eqD3}
&A(t)=\Omega_{d}\left[1-\cos\left(\frac{2\pi t}{t_{g}}\right)\right].
\end{aligned}
\end{equation}
where $t_{g}$ is the gate length (excluding the coupler bias ramp time) and $\Omega_{d}$ is the 
peak drive amplitude. 

For CZ gate calibration, the gate parameters, i.e., the drive peak amplitude $\Omega_{d}$ and the drive frequency $\omega_{d}$, are optimized by minimizing leakage out of the computational subspace and the conditional phase error. To evaluate the intrinsic gate performance without considering decoherence, we use the state-averaged gate fidelity (up to single-qubit Z phases) given by~\cite{Pedersen2007}
\begin{equation}
\begin{aligned}\label{eqD4}
F=\frac{{\rm Tr}(\tilde{U}^{\dagger}\tilde{U})+|{\rm Tr}(U_{\rm CZ}^{\dag}\tilde{U})|^{2}}{N(N+1)},
\end{aligned}
\end{equation}
where $\tilde{U}$ denotes the actual evolution operator truncated to the two-qubit computational subspace spanned by $\{|00\rangle,|01\rangle,|10\rangle,|11\rangle\}$, $U_{\rm cz}$ denotes the ideal CZ gate, and $N=4$ is the dimension of the computational subspace.

\begin{table}[!htb]
\caption{\label{tab:QC_bias_parameters} Coupler bias parameters (Coupler bias), conditional frequency shift (Conditional shift) $\delta\omega_{12,i}$, aand spectator-induced shift in the gate transition (Spectator-induced shift ) $\delta\omega_{|11\rangle\leftrightarrow|21\rangle}$ for the quarton-circuit coupler used in the gate performance analysis shown in Fig.~\ref{fig3} of the main text.}
\begin{ruledtabular}
\begin{tabular}{cccc}
Configuration                               & $Q_{1}$-$Q_{2}$  & $S_1$-$Q_{1}$-$Q_{2}$ & $Q_{1}$-$Q_{2}$-$S_2$ \\\hline
Coupler bias ($\Phi_{\text{ext},c}/\Phi_0$) & [0.456]              & [0.140,0.456]             & [0.456,0] \\
Conditional shift (MHz)                     & 54.711           & 54.851                & 51.728  \\
Spectator-induced shift (kHz)               & -                & $<1$              & $<1$
\end{tabular}
\end{ruledtabular}
\end{table}

\begin{table}[!htb]
\caption{\label{tab:FC_bias_parameters} Coupler bias parameters (Coupler bias), conditional frequency shift (Conditional shift) $\delta\omega_{12,i}$, aand spectator-induced shift in the gate transition (Spectator-induced shift ) $\delta\omega_{|11\rangle\leftrightarrow|21\rangle}$ for the fluxonium-circuit coupler used in the gate performance analysis shown in Fig.~\ref{fig3} of the main text.}
\begin{ruledtabular}
\begin{tabular}{cccc}
Configuration                               & $Q_{1}$-$Q_{2}$  & $S_1$-$Q_{1}$-$Q_{2}$ & $Q_{1}$-$Q_{2}$-$S_2$ \\\hline
Coupler bias ($\Phi_{\text{ext},c}/\Phi_0$) & [0]              & [0.295,0]             & [0.080,0.283]  \\
Conditional shift (MHz)                     & 42.789           & 41.494                & 39.759 \\
Spectator-induced shift (kHz)               & -                & 82              & 72
\end{tabular}
\end{ruledtabular}
\end{table}

To characterize the spectator-induced gate error in the studied architecture, Figs.~\ref{fig3}(d) and~\ref{fig3}(h) of the main text show the gate error for CZ gates applied to $Q_{1}$ and $Q_{2}$, taking into account the presence of spectator $S_{1}$ (coupled to $Q_{1}$) or spectator $S_{2}$ (coupled to $Q_{2}$). Following typical protocols~\cite{Krinner2020,Cai2021,Zhao2023}, the CZ gate is calibrated with the spectator qubit prepared in $|0\rangle$ and its performance is then evaluated for both $|0\rangle$ and $|1\rangle$ states of 
the spectator, using Eq.~(\ref{eqD4}). 

For easy reference, we summarize in Table~\ref{tab:QC_bias_parameters} the coupler bias parameters, the resulting conditional frequency shift $\delta\omega_{12,i}$, and the spectator-induced shift in the gate transition $\delta\omega_{|11\rangle\leftrightarrow|21\rangle}$ for the quarton-circuit coupler, and in Table~\ref{tab:FC_bias_parameters} the corresponding quantities for the fluxonium-circuit 
coupler, as relevant to the gate performance analysis presented in Fig.~\ref{fig3} of the main 
text.



\begin{thebibliography}{99}

\bibitem{Jiang2025} Y.-Y. Jiang, C. Deng, H. Fan, B.-Y. Li, L. Sun, X.-S. Tan, W. Wang, G.-M. Xue, F. Yan, H.-F. Yu, Y.-S. Zhang, Y.-R. Zhang, and C.-L. Zou, Advancements in superconducting quantum computing, \href{https://doi.org/10.1093/nsr/nwaf246}{Natl. Sci. Rev. \textbf{12}, nwaf246 (2025)}.

\bibitem{Manucharyan2009} V. E. Manucharyan, J. Koch, L. I. Glazman, and M. H. Devoret, Fluxonium: Single Cooper-pair circuit free of charge offsets, \href{https://doi.org/10.1126/science.1175552}{Science \textbf{326}, 113 (2009)}.

\bibitem{Nguyen2019} L. B. Nguyen, Y.-H. Lin, A. Somoroff, R. Mencia, N. Grabon, and V. E. Manucharyan, High-coherence fluxonium qubit, \href{https://doi.org/10.1103/PhysRevX.9.041041}{Phys. Rev. X \textbf{9}, 041041 (2019)}.

\bibitem{Somoroff2023} A. Somoroff, Q. Ficheux, R. A. Mencia, H. Xiong, R. Kuzmin, and V. E. Manucharyan, Millisecond coherence in a superconducting qubit, \href{https://doi.org/10.1103/PhysRevLett.130.267001}{Phys. Rev. Lett. \textbf{130}, 267001 (2023)}.

\bibitem{Rower2024} D. A. Rower, L. Ding, H. Zhang, M. Hays, J. An, P. M. Harrington, I. T. Rosen, J. M. Gertler, T. M. Hazard, B. M. Niedzielski, M. E. Schwartz, S. Gustavsson, K. Sernia, J. A. Grover, and W. D. Oliver, Suppressing Counter-Rotating Errors for Fast Single-Qubit Gates with
 Fluxonium, \href{http://dx.doi.org/10.1103/PRXQuantum.5.040342}{PRX Quantum \textbf{5}, 040342 (2024)}.

\bibitem{Ding2023} L. Ding, M. Hays, Y. Sung, B. Kannan, J. An, A. Di Paolo, A. H. Karamlou, T. M. Hazard, K. Azar, D. K. Kim, B. M. Niedzielski, A. Melville, M. E. Schwartz, J. L. Yoder, T. P. Orlando, S. Gustavsson, J. A. Grover, K. Serniak, and W. D. Oliver, High-fidelity, frequency-flexible two-qubit fluxonium gates with a transmon coupler, \href{https://doi.org/10.1103/PhysRevX.13.031035}{Phys. Rev. X \textbf{13}, 031035 (2023)}.

\bibitem{Zhang2024} H. Zhang, C. Ding, D. Weiss, Z. Huang, Y. Ma, C. Guinn, S. Sussman, S. P. Chitta, D. Chen, A. A. Houck, J. Koch, and D. I. Schuster, Tunable inductive coupler for high-fidelity gates between fluxonium qubits, \href{https://doi.org/10.1103/PRXQuantum.5.020326}{PRX Quantum \textbf{5}, 020326 (2024)}.
    
\bibitem{Lin2025} W.-J. Lin, H. Cho, Y. Chen, M. G. Vavilov, C. Wang, and V. E. Manucharyan, 24 Days-Stable CNOT Gate on Fluxonium Qubits with Over 99.9$\%$ Fidelity, \href{https://doi.org/10.1103/PRXQuantum.6.010349}{PRX Quantum \textbf{6}, 010349 (2025)}.  

\bibitem{Koch2007} J. Koch, T. M. Yu, J. Gambetta, A. A. Houck, D. I. Schuster, J. Majer, A. Blais, M. H. Devoret, S.
    M. Girvin, and R. J. Schoelkopf, Charge-insensitive qubit design derived from the cooper pair box, \href{https://doi.org/10.1103/PhysRevA.76.042319}{Phys. Rev. A \textbf{76}, 042319 (2007)}.

\bibitem{Zhao2026} P. Zhao, G. Zhao, S. Li, C. Zha, and M. Gong, Scalable fluxonium qubit architecture with tunable interactions between non-computational levels, \href{https://link.aps.org/doi/10.1103/mfjl-5rk6}{Phys. Rev. Appl. \textbf{25}, 044072 (2026)}.

\bibitem{Zhao2026a} P. Zhao, P. Xu, and Z.-Y. Xue, Scalable native multiqubit gates via engineered noncomputational-state interactions in superconducting fluxonium qubits, \href{https://doi.org/10.1103/xwny-vkft}{Phys. Rev. A \textbf{113}, 022604 (2026)}.

\bibitem{Zwanenburg2026} M. F. S. Zwanenburg and C. K. Andersen, Crosstalk in Multi-Qubit Fluxonium Architectures with Transmon Couplers, \href{https://doi.org/10.48550/arXiv.2603.09870}{arXiv:2603.09870}.
    
\bibitem{Zhan2026} Z. Zhan, Z. Li, F. Wang, W. Lan, Xi. Pan, L. Xiang, X. Dou, R. Gao, G. Gong, Y. Guo \emph{et al.}, Scalable Fluxonium Quantum Processors via Tunable-Coupler Architecture, \href{https://arxiv.org/abs/2604.13363}{arXiv:2604.13363}.      
   
\bibitem{Chan2026} G. X. Chan, W. Lan, T. Wang, X. Ma, C. Deng, and L. Jin, System-Level Design of Scalable Fluxonium Quantum Processors with Double-Transmon Couplers, \href{https://doi.org/10.48550/arXiv.2604.26373}{arXiv:2604.26373}.   


\bibitem{Nesterov2018} K. N. Nesterov, I. V. Pechenezhskiy, C. Wang, V. E. Manucharyan, and M. G. Vavilov, Microwave-activated controlled-$Z$ gate for fixed-frequency fluxonium qubits, \href{https://doi.org/10.1103/PhysRevA.98.030301}{Phys. Rev. A \textbf{98}, 030301 (2018)}.
    
\bibitem{Nguyen2022} L. B. Nguyen, G. Koolstra, Y. Kim, A. Morvan, T. Chistolini, S. Singh, K. N. Nesterov, C. Jünger, L. Chen, Z. Pedramrazi, B. K. Mitchell, J. M. Kreikebaum, S. Puri, D. I. Santiago, and I Siddiqi, Blueprint for a High-Performance Fluxonium Quantum Processor, \href{https://doi.org/10.1103/PRXQuantum.3.037001}{PRX Quantum \textbf{3}, 037001 (2022)}.       
    
\bibitem{Moreno2025} V. D. Moreno, N. D. Dimitrov, V. E. Manucharyan, and M. G. Vavilov, CNOT gates in inductively coupled multi-fluxonium systems, \href{https://doi.org/10.48550/arXiv.2512.11756}{arXiv:2512.11756}.

\bibitem{Kugut2025} A. A. Kugut, G. S. Mazhorin, and I. A. Simakov, Interaction-Resilient Scalable Fluxonium Architecture with All-Microwave Gates, \href{https://arxiv.org/abs/2512.21189}{arXiv:2512.21189}.

\bibitem{Huang2026} E. Y. Huang and C. K. Andersen, Exploration of Fluxonium Parameters for Capacitive Cross-Resonance Gates, \href{https://arxiv.org/abs/2603.17936}{arXiv:2603.17936}.    

\bibitem{Brink2018} M. Brink, J. M. Chow, J. Hertzberg, E. Magesan, and Sami Rosenblatt, Device challenges for near term superconducting quantum processors: frequency collisions, \href{https://doi.org/10.1109/IEDM.2018.8614500}{2018 IEEE Int. Electron Devices Meeting (IEDM), (2018)}.

\bibitem{Malekakhlagh2020}  M. Malekakhlagh, E. Magesan, and D. C. McKay, First-principles analysis of cross-resonance gate operation, \href{https://doi.org/10.1103/PhysRevA.102.042605}{Phys. Rev. A \textbf{102}, 042605 (2020)}.

\bibitem{Zhao2023} P. Zhao, Mitigation of quantum crosstalk in cross-resonance-based qubit architectures, \href{http://dx.doi.org/10.1103/PhysRevApplied.20.054033}{Phys. Rev. Appl. \textbf{20}, 054033 (2023)}.

\bibitem{Heya2024} K. Heya, M. Malekakhlagh, S. Merkel, N. Kanazawa, and E. Pritchett, Floquet analysis of frequency collisions, \href{http://dx.doi.org/10.1103/PhysRevApplied.21.024035}{Phys. Rev. Appl. \textbf{21}, 024035 (2024)}.
    
\bibitem{Moskalenko2021}  I. N. Moskalenko, I. S. Besedin, I. A. Simakov, and A. V. Ustinov, Tunable coupling scheme for implementing two-qubit gates on fluxonium qubits, \href{https://doi.org/10.1063/5.0064800}{Appl. Phys. Lett. \textbf{119}, 194001 (2021)}.

\bibitem{Moskalenko2022}  I. N. Moskalenko, I. A. Simakov, N. N. Abramov, A. A. Grigorev, D. O. Moskalev, A. A. Pishchimova, N. S. Smirnov, E. V. Zikiy, I. A. Rodionov, and I. S. Besedin, High fidelity two-qubit gates on fluxoniums using a tunable coupler, \href{https://doi.org/https://doi.org/10.1038/s41534-022-00644-x}{npj Quantum Inf. \textbf{8}, 130 (2022)}.
    
\bibitem{Chen2014}  Y. Chen, C. Neill, P. Roushan, N. Leung, M. Fang, R. Barends, J. Kelly, B. Campbell, Z. Chen, and B. Chiaro \emph{et al.}, Qubit architecture with high coherence and fast tunable coupling, \href{https://doi.org/10.1103/PhysRevLett.113.220502}{Phys. Rev. Lett. \textbf{113}, 220502 (2014)}.    
    
\bibitem{Yan2018} F. Yan, P. Krantz, Y. Sung, M. Kjaergaard, D. L. Campbell, T. P. Orlando, S. Gustavsson, and W. D. Oliver, Tunable Coupling Scheme for Implementing High-Fidelity Two-Qubit Gates, \href{https://doi.org/10.1103/PhysRevApplied.10.054062}{Phys. Rev. Appl. \textbf{10}, 054062 (2018)}.

\bibitem{Goto2022} H. Goto, Double-Transmon Coupler: Fast Two-Qubit Gate with No Residual Coupling for Highly Detuned Superconducting Qubits, \href{https://doi.org/10.1103/PhysRevApplied.18.034038}{Phys. Rev. Appl. \textbf{18}, 034038 (2022)}.

\bibitem{Campbell2023} D. L. Campbell, A.a Kamal, L. Ranzani, M. Senatore, and M. D. LaHaye, Modular Tunable Coupler for Superconducting Circuits, \href{http://dx.doi.org/10.1103/PhysRevApplied.19.064043}{Phys. Rev. Appl. \textbf{19}, 064043 (2023)}.

\bibitem{Miyanaga2021} T. Miyanaga, A. Tomonaga, H. Ito, H. Mukai, and J.S. Tsai, Ultrastrong Tunable Coupler Between Superconducting LC Resonators, \href{http://dx.doi.org/10.1103/PhysRevApplied.16.064041}{Phys. Rev. Appl. \textbf{16}, 064041 (2021)}. 
    
\bibitem{Zhao2026b} P. Zhao, P. Xu, and Z.-Y. Xue, Long-range tunable coupler for modular fluxonium quantum processors, \href{https://doi.org/10.48550/arXiv.2509.04762}{arXiv:2604.12261}.     
    
\bibitem{Rosenfeld2024} E. L. Rosenfeld, C. T. Hann, D. I. Schuster, M. H. Matheny, and A. A. Clerk, High-Fidelity Two-Qubit Gates between Fluxonium Qubits with a Resonator Coupler, \href{http://dx.doi.org/10.1103/PRXQuantum.5.040317}{PRX Quantum \textbf{5}, 040317 (2024)}.    

\bibitem{Yan2020} F. Yan, Y. Sung, P. Krantz, A. Kamal, D. K. Kim, J. L. Yoder, T. P. Orlando, S. Gustavsson, and W. D. Oliver, Engineering Framework for Optimizing Superconducting Qubit Designs, \href{https://doi.org/10.48550/arXiv.2006.04130}{arXiv:2006.04130}.
    
\bibitem{Ficheux2021} Q. Ficheux, L. B. Nguyen, A. Somoroff, H. Xiong, K. N. Nesterov, M. G. Vavilov, and V. E. Manucharyan, Fast logic with slow qubits: Microwave-activated controlled-Z gate on low-frequency fluxoniums, \href{https://doi.org/10.1103/PhysRevX.11.021026}{Phys. Rev. X \textbf{11}, 021026 (2021)}.    
    
\bibitem{Simakov2023} I. A. Simakov, G. S. Mazhorin, I. N. Moskalenko, N. N. Abramov, A. A. Grigorev, D. O. Moskalev, A. A. Pishchimova, N. S. Smirnov, E. V. Zikiy, I. A. Rodionov, and I. S. Besedin, Coupler Microwave-Activated Controlled-Phase Gate on Fluxonium Qubits, \href{http://dx.doi.org/10.1103/PRXQuantum.4.040321}{PRX Quantum \textbf{4}, 040321 (2023)}.    
    
\bibitem{Zhao2025c} P. Zhao, P. Xu, and Z.-Y. Xue, Fast entangling gates on fluxoniums via parametric modulation of plasmon interaction, \href{https://doi.org/10.48550/arXiv.2509.04762}{arXiv:2509.04762}.    
    
\bibitem{Chow2013} J. M. Chow, J. M. Gambetta, A. W. Cross, S. T. Merkel, C. Rigetti, and M. Steffen, Microwave-activated conditional-phase gate for superconducting qubits, \href{https://doi.org/10.1088/1367-2630/15/11/115012}{New J. Phys. \textbf{15}, 115012 (2013)}.    

\bibitem{Niskanen2007} A. O. Niskanen, K. Harrabi, F. Yoshihara, Y. Nakamura, S. Lloyd, and J. S. Tsai, Quantum coherent tunable coupling of superconducting qubits, \href{https://doi.org/10.1126/science.1141324}{Science \textbf{316}, 723 (2007)}.

\bibitem{Mundada2019} P. Mundada, G. Zhang, T. Hazard, and A. Houck, Suppression of Qubit Crosstalk in a Tunable Coupling Superconducting Circuit, \href{http://dx.doi.org/10.1103/PhysRevApplied.12.054023}{Phys. Rev. Appl. \textbf{12}, 054023 (2019)}.

\bibitem{DiCarlo2009} L. DiCarlo, J. M. Chow, J. M. Gambetta, L. S. Bishop, B. R. Johnson, D. I. Schuster, J. Majer, A. Blais, L. Frunzio, S. M. Girvin, and R. J. Schoelkopf, Demonstration of two-qubit algorithms with a superconducting quantum processor, \href{https://doi.org/10.1038/nature08121}{Nature (London) \textbf{460}, 240 (2009)}.

\bibitem{Krinner2020} S. Krinner, S. Lazar, A. Remm, C.K. Andersen, N. Lacroix, G. J. Norris, C. Hellings, M. Gabureac, C. Eichler, and A. Wallraff, Benchmarking Coherent Errors in Controlled-Phase Gates due to Spectator Qubits, \href{http://dx.doi.org/10.1103/PhysRevApplied.14.024042}{Phys. Rev. Appl. \textbf{14}, 024042 (2020)}.

\bibitem{Cai2021} T.-Q. Cai, X.-Y. Han, Y.-K. Wu, Y.-L. Ma, J.-H. Wang, Z.-L. Wang, H.-Y. Zhang, H.-Y. Wang, Y.-P. Song, and L.-M. Duan, Impact of Spectators on a Two-Qubit Gate in a Tunable Coupling Superconducting Circuit, \href{https://doi.org/10.1103/PhysRevLett.127.060505}{Phys. Rev. Lett. \textbf{127}, 060505 (2021)}.

\bibitem{Motzoi2009} F. Motzoi, J. M. Gambetta, P. Rebentrost, and F. K. Wilhelm, Simple Pulses for Elimination of Leakage in Weakly Nonlinear Qubits, \href{https://doi.org/10.1103/PhysRevLett.103.110501}{Phys. Rev. Lett. \textbf{103}, 110501 (2009)}.

\bibitem{Pedersen2007} L. H. Pedersen, N. M. M{\o}ller, and K. M{\o}lmer, Fidelity of quantum operations,
    \href{https://doi.org/10.1016/j.physleta.2007.02.069}{Phys. Lett. A \textbf{367}, 47 (2007)}.

\bibitem{Breuckmann2021} N. P. Breuckmann and J. N. Eberhardt, Quantum Low-Density Parity-Check Codes, \href{https://doi.org/10.1103/PRXQuantum.2.040101}{PRX Quantum \textbf{2}, 040101 (2021)}. 

\bibitem{Bravyi2024} S. Bravyi, A. W. Cross, J. M. Gambetta, D. Maslov, P. Rall, and T. J. Yoder, High-threshold and low-overhead fault-tolerant quantum memory, \href{https://doi.org/10.1038/s41586-024-07107-7}{Nature \textbf{627}, 778 (2024)}.

\bibitem{Marxer2023} F. Marxer, A. Veps\"{a}l\"{a}inen, S. W. Jolin, J. Tuorila, A. Landra, C. Ockeloen-Korppi, W. Liu, O. Ahonen, A. Auer \emph{et al.}, Long-Distance Transmon Coupler with CZ-Gate Fidelity above 99.8$\%$, \href{http://dx.doi.org/10.1103/PRXQuantum.4.010314}{PRX Quantum \textbf{4}, 010314 (2023)}.

\bibitem{Liang2023} G.-H. Liang, X.-H. Song, C.-L. Deng, X.-Y. Gu, Y. Yan, Z.-Y. Mei, S.-L. Zhao, Y.-Z. Bu, Y.-X. Xiao \emph{et al.}, Tunable-coupling architectures with capacitively connecting pads for large-scale superconducting multiqubit processors, \href{https://doi.org/10.1103/PhysRevApplied.20.044028}{Phys. Rev. Appl. \textbf{20}, 044028 (2023)}.

\bibitem{Liu2023} F.-M. Liu, C. Wang, M.-C. Chen, H. Chen, S.-W. Li, Z.-X. Shang, C. Ying, J.-W. Wang, Y.-H. Huo, C.-Z. Peng, X. Zhu, C.-Y. Lu, and J.-W. Pan, Quantum computer-aided design for advanced superconducting qubit: Plasmonium, \href{https://doi.org/10.1016/j.scib.2023.06.030}{Sci. Bull. \textbf{68}, 1625 (2023)}.

\bibitem{Azar2026} K. Azar, L. Ateshian, M. T. Randeria, R. DePencier Pi\~{n}ero, J. M. Gertler, J. An, F. Contipelli, L. Ding, M. Gingras, K. Grossklaus \emph{et al.}, Characterization and Comparison of Energy Relaxation in Fluxonium Qubits, \href{https://arxiv.org/abs/2603.23636}{arXiv:2603.23636}.    

\bibitem{Abad2025} T. Abad, Y. Schattner, A. F. Kockum, and G. Johansson, Impact of decoherence on the fidelity of quantum gates leaving the computational subspace, \href{https://doi.org/10.22331/q-2025-04-03-1684}{Quantum \textbf{9}, 1684 (2025)}.

\bibitem{Gambetta2016} J. M. Gambetta, C. E. Murray, Y.-K.-K. Fung, D. T. McClure, O. Dial, W. Shanks, J. W. Sleight, and M. Steffen, Investigating surface loss effects in superconducting transmon qubits, \href{https://doi.org/10.1109/TASC.2016.2629670}{IEEE Trans. Appl. Supercond. \textbf{27}, 1 (2016)}.

\bibitem{Zajac2021} D. M. Zajac, J. Stehlik, D. L. Underwood, T. Phung, J. Blair, S. Carnevale, D. Klaus, G. A. Keefe, A. Carniol, M. Kumph, M. Steffen, and O. E. Dial, Spectator Errors in Tunable Coupling Architectures, \href{https://doi.org/10.48550/arXiv.2108.11221}{arXiv:2108.11221}.

\bibitem{Klimov2024} P. V. Klimov, A. Bengtsson, C. Quintana, A. Bourassa, S. Hong, A. Dunsworth, K. J. Satzinger, W. P. Livingston, V. Sivak, M. Y. Niu \emph{et al.}, Optimizing quantum gates towards the scale of logical qubits, \href{https://doi.org/10.1038/s41467-024-46623-y}{Nat. Commun. \textbf{15}, 2442 (2024)}.  

\bibitem{McKay2017} D. C. McKay, C. J. Wood, S. Sheldon, J. M. Chow, and J. M. Gambetta, Efficient Z gates for quantum computing, \href{http://dx.doi.org/10.1103/PhysRevA.96.022330}{Phys. Rev. A \textbf{96}, 022330 (2017)}.


\end{thebibliography}
\end{document}